\documentclass[11pt,preprint]{aastex}

\newcommand{\kms}{km~s$^{-1}$}
\newcommand{\hii}{H II~}

\begin{document}

\slugcomment{Accepted for publication in the Astrophysical Journal}

\title{Oxygen and Nitrogen in Isolated Dwarf Irregular Galaxies}

\author{Liese van Zee}
\affil{Astronomy Department, Indiana University, 727 E 3rd St, Bloomington, IN 47405}
\email{vanzee@astro.indiana.edu}
\and
\author{Martha P. Haynes}
\affil{Center for Radiophysics and Space Research and
National Astronomy and Ionosphere Center,\altaffilmark{1} Cornell University, Ithaca, NY 14853}
\email{haynes@astro.cornell.edu}

\altaffiltext{1}{The National Astronomy and Ionosphere Center is
operated by Cornell University under a cooperative agreement with the
National Science Foundation.}

\begin{abstract}
We present long slit optical spectroscopy of 67 \hii regions in 21 dwarf irregular galaxies
to investigate the enrichment of oxygen, nitrogen, neon, sulfur, and argon in
low mass galaxies.  Oxygen abundances are obtained via direct detection of the temperature
sensitive emission lines for 25 \hii regions; for the remainder of the sample,
oxygen abundances are estimated from strong line calibrations.  The
direct abundance determinations are compared to the strong-line abundance
calibrations of both McGaugh (1991) and Pilyugin (2000).  While the McGaugh (1991)
calibration yields a statistical offset of 0.07 dex, the photoionization model
grid traces the appropriate iso-metallicity contour shape in the
R23-O32 diagnostic diagram.  In contrast, while the Pilyugin (2000) calibration 
yields a negligible statistical offset, the residuals in this strong-line 
calibration method are correlated with ionization parameter.  Thus, these observations
indicate that oxygen abundances will be overestimated by the p-method for \hii regions 
with low ionization parameters.  Global oxygen and nitrogen abundances for
this sample of dwarf irregular galaxies are examined in the context of open and
closed box chemical evolution models.  While several galaxies are consistent
with closed box chemical evolution, the majority of this sample have an 
effective yield $\sim$1/4 of the expected yield for a constant star formation
rate and Salpeter IMF, indicating 
that either outflow of enriched gas or inflow of pristine gas has occurred.  
The effective yield strongly correlates with M$_{\rm H}$/L$_{\rm B}$ in
the sense that gas-rich galaxies are more likely to be closed systems.
However, the effective yield does not appear to correlate with other
global parameters such as dynamical mass, absolute magnitude, star formation
rate or surface brightness.  In addition, open and closed systems are not 
identified easily in other global abundance measures; for example, the observed 
correlation between luminosity and metallicity is consistent with other recent 
results in the literature.  A correlation is found between the observed
nitrogen--to--oxygen ratio and the color of the underlying stellar population;
redder dwarf irregular galaxies have higher N/O ratios than blue dwarf irregular
galaxies.  The relative abundance ratios are interpreted in the context of delayed release
of nitrogen and varied star formation histories.
\end{abstract}


\section{Introduction}
Gas phase oxygen and nitrogen abundances are two of the most robust
measures of the intrinsic metallicity of gas-rich low mass galaxies.
Over the last three decades, numerous studies have measured
the oxygen and nitrogen abundances of \hii regions in dwarf 
irregular galaxies \citep[e.g.,][]{ACJV79,LPRST79,KD81,CTM86,G90,MH96,vHS97a,IT98,HGO02,SCM03,LGH03,LSM04}.
\citet{LPRST79} were the first to note a strong correlation between 
metallicity and mass, in the sense that more massive galaxies also
have higher measured oxygen abundances \citep[see also][]{KD81}; 
\citet{SKH89} determined that a stronger correlation was found between 
metallicity and luminosity \citep[see also][for an alternative view, 
see Hidalgo-G\'amez \& Olofsson 1998]{RM95,HH99,P01,T04}.
However, despite extensive observational effort, the origin of the luminosity--metallicity 
relation is still poorly understood.  One possibility is that it 
represents an evolutionary sequence, in which more luminous 
galaxies have processed a larger fraction of their raw materials.  
Alternatively, it may represent a mass retention sequence, in which more
luminous (massive) galaxies are able to retain a larger fraction
of their processed materials.   In the former case, one would
expect to see a stronger correlation between gas-mass fraction
and metallicity; in the latter case, one would expect to see
a difference between the metallicity--luminosity relation
traced by elements formed in highly energetic Type II SN (such
as oxygen) and in the relation traced by elements formed in
lower mass stars (such as nitrogen).
Clearly, one key to understanding the metallicity--luminosity
relation is to look for additional parameters which are responsible for 
the scatter in the correlation.  With this goal in mind, we present 
oxygen and nitrogen abundances in 67 \hii regions in 21 well--studied
dwarf irregular galaxies which span a moderate range in luminosity.

Of particular interest is whether there are significant second parameter 
effects in the global abundance trends due to variations of surface brightness,
color, gas mass fraction, average star formation rate, or other
evolutionary effects \citep[see, e.g.,][]{G02,Ke03,T04}.  
The galaxies in this study are part of an extensive 
optical imaging program of isolated dwarf irregular galaxies and thus a substantial
amount of supporting data exists in the literature \citep{vHS97a,vZ00}. 
This sample includes galaxies with a wide range of surface brightness 
(20 $< \mu_B^0 <$ 26 mag arcsec$^{-2}$), gas mass fraction 
(0.4 $< f_{\rm gas} <$ 0.95), and current star 
formation rate (0.002 $< SFR <$ 0.16 M$_{\odot}$ yr$^{-1}$), 
but only a modest range of color (B-V $\sim$ 0.4 $\pm$ 0.1).
Most of the galaxies in this sample are relatively isolated
(d $>$ 200 kpc to nearest neighbor), and thus  have not had 
significant alterations of their star formation activity or
elemental enrichment due to tidal interactions within the
last few Gyr.  Thus, elemental abundance measurements of
this well--studied sample of galaxies will provide 
insight into the chemical evolution and enrichment processes
of low mass galaxies.

This paper presents the results of long--slit optical spectroscopy of
21 dwarf irregular galaxies.  The observations and data reduction 
procedures are discussed in Section 2.  Derivation of nebular abundances 
via direct and strong line methods are described in Section 3.  Correlations 
between both oxygen and nitrogen abundances and other global parameters
are discussed in the context of open and closed box chemical
evolution models in Section 4.  Section 5 contains a brief summary
of the conclusions.

\section{Observations}

\subsection{Sample Selection}
The dwarf irregular galaxies for the present study were selected from 
the optical imaging sample of \citet{vZ00,vZ01}.  This sample of dwarf 
irregular galaxies was selected primarily from the 
{\em Uppsala General Catalog of Galaxies} \citep[UGC]{UGC}
based on morphological classification, apparent diameter (D $<$ 7\arcmin), 
gas--richness ($\int S dv$/W $>$ 50 mJy), and apparent isolation (no known
neighbor within 30\arcmin~and 500 \kms). 
The present sample contains galaxies classified as
late--type spiral or irregular, with M$_{B} > -18.0$,
and with heliocentric velocities less than 2000 \kms.
Basic parameters for the selected galaxies are tabulated 
in Table \ref{tab:gals}.

Table \ref{tab:gals} includes a distance estimate
adopted from the literature for those galaxies with
TRGB observations [UGC 685 \citep{MA02}, 
HKK97 L14 \citep{Ke04},
UGC 4483 \citep{U4483},
UGC 8651 \citep{Ke02},
and UGC 9240 \citep{Ke02}];  for those galaxies without TRGB distances,
the distance to each system is calculated
from the systemic velocity using a Virgocentric infall model and
an H$_0$ of 75 km s$^{-1}$ Mpc$^{-1}$.   Optical colors,
sizes, luminosities, M$_H$/L$_B$, central surface brightnesses, and dynamical
mass estimates listed in Table \ref{tab:gals}
are taken from the tabulation of \citet{vZ00}.  The absolute blue 
magnitudes and colors have been corrected for Galactic extinction, 
but not for internal extinction or for nebular contributions to 
the broadband luminosity.  Finally, the oxygen and nitrogen abundances
tabulated in Table 1 are the mean values from the present study.

\subsection{Optical Spectroscopy}
Optical spectroscopic observations of \hii regions in 21 dwarf irregular
galaxies were obtained with the Double Spectrograph on the 5m Palomar\footnote{Observations 
at the Palomar Observatory were made as part of a continuing cooperative
agreement between Cornell University and the California Institute of
Technology.} telescope during several observing runs between 1999 January and 2002 April. 
The  observations were obtained in long--slit mode; the 2\arcmin~long--slit
was set to an aperture 2\arcsec~wide.  The two sides of the spectrograph
were set to complementary spectral resolution, and provided continuous
wavelength coverage between 3600 \AA~and 7600 \AA.  The blue side was
equipped with a 600 l/mm grating (spectral resolution of 5.0 \AA~or 
1.72 \AA~pix$^{-1}$); the red side was equipped with a 300 l/mm grating 
(spectral resolution of 7.9 \AA~or 2.47 \AA~pix$^{-1}$).  
We also include observations of CGCG 007-025, which were obtained
in 1997 January, during an observing run described in \citet{vSHOB98}.

The H$\alpha$ images of \citet{vZ00}  were used to determine the
appropriate slit positions and position angles (except for UGC 4483,
where an H$\alpha$ image was kindly provided by Evan Skillman).  
Astrometric plate solutions were calculated using the coordinates
of the bright stars in the APM catalog \citep{APM}, yielding
positions accurate to within 1\arcsec.  In all cases, the telescope
was centered on a nearby star, and then moved to the \hii region via
blind offsets.  The slit positions and position angles are tabulated
in Table \ref{tab:slits}.  The slit was set to a position angle 
close to the parallactic angle at the time of the observations.  
The observations were scheduled to maximize
the number of \hii regions that could be observed simultaneously.
In some instances, faint \hii regions were given higher priority if the
parallactic angle was favorable for observations of multiple \hii regions
in a single slit position.  In all, the 31 pointings yielded high 
signal--to--noise observations of 67 \hii regions.
Throughout this paper, the H II region
nomenclature is based on east--west and north--south offsets from
the galaxy center.  These offsets are derived either from the 
pointing center (if only one H II region was in the slit), or 
are computed for each H II region from the pointing center, 
the slit position angle, and the spatial scale of the spectrum. 

\begin{figure}
\plotone{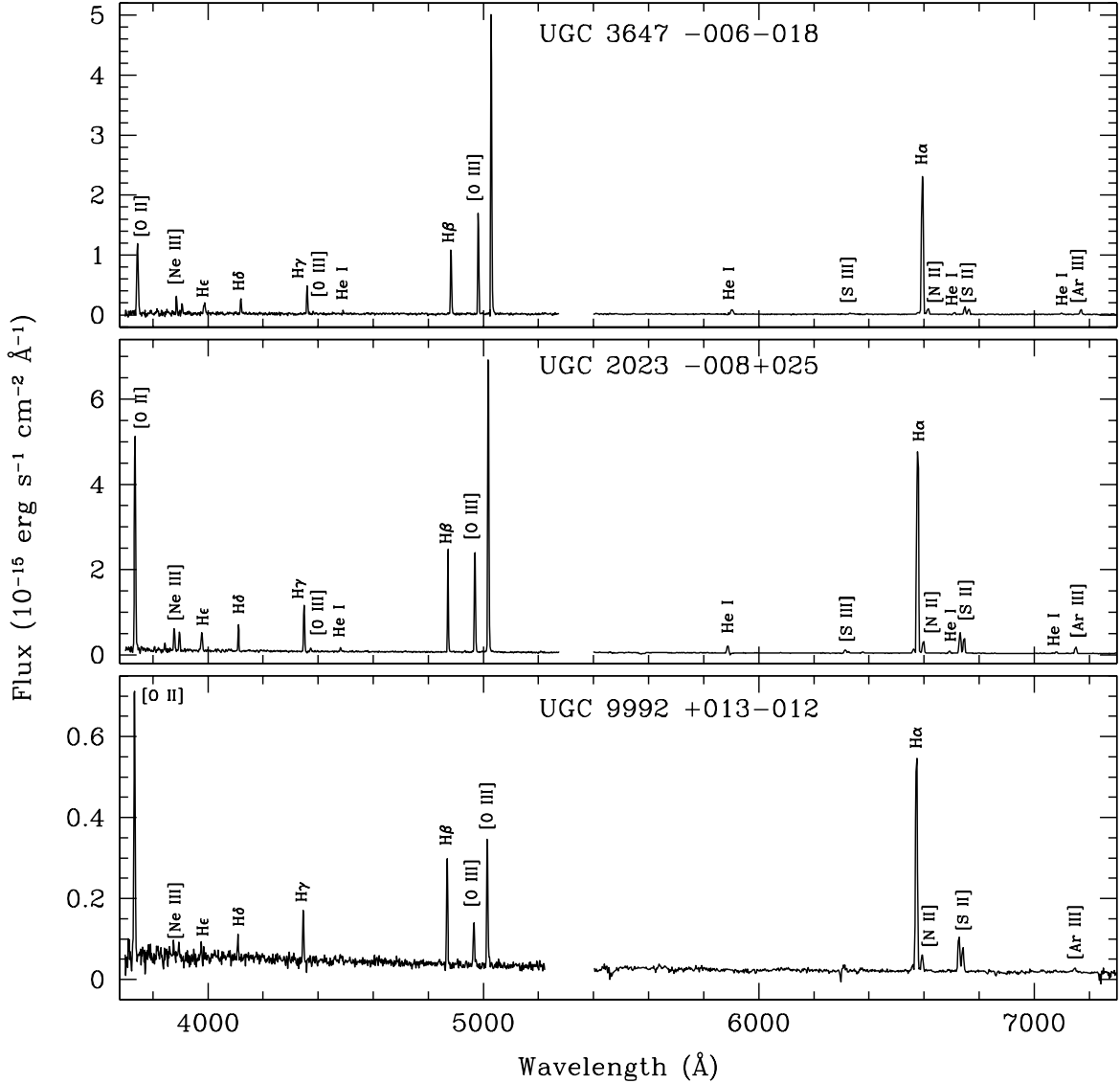}
\caption[]{Representative optical spectra 
of H II regions in three low metallicity galaxies.    All 
3 H II regions shown here have similar oxygen abundances (12 + log(O/H) = 8.0).
However, the relative line strengths of [O II] and [O III] indicate
that these 3 H II regions have very different ionization parameters.
UGC 3647 -006-018 has one of the highest ionization parameters 
(\={U} $\sim 0.005$) in this sample, while
UGC 2023 -008+025 is representative of the typical H II region
in this sample (\={U} $\sim 0.002$) and UGC 9992 +013-012 has
one of the lowest ionization parameters (\={U} $\sim 0.0006$).
\label{fig:spec} }
\end{figure}

The long--slit optical spectra were processed using standard methods in
the IRAF\footnote{IRAF is distributed by the National Optical Astronomy 
Observatories.} package.  The flat field was created from a combination
of dome flats and twilight flats to correct for the slit illumination and
wavelength responsivity.  Wavelength rectification was based on arclamps
taken before and after each series of observations.  The images were rectified
in the spatial dimension based on the trace of stars at different positions
along the slit.  One-dimensional (1-d) spectra were then extracted from the two 
dimensional images.  Aperture sizes for the extracted 1-d spectra were 
matched to the observations: faint \hii regions 
(isolated \hii regions) were extracted with apertures defined at 10\% of the peak, 
while brighter \hii regions (\hii region complexes) were
extracted with both wide (10\% of the peak) and narrow (3\arcsec) apertures.
In the final analysis, the narrow (3\arcsec) apertures were favored for the
bright \hii regions since they had less contamination from blending.
The one-dimensional spectra were flux calibrated using standard stars
from the compilation of \citet{O90}.  Typical rms for the flux calibration
is 0.015.  While the nights were typically non-photometric, the relative 
line ratios should be robust.  Representative spectra are shown in 
Figure \ref{fig:spec}.  The excellent agreement in the
continuum levels in the blue and red spectra illustrated
 in Figure \ref{fig:spec} confirms that
the extraction regions were well matched in the two cameras.

\section{Nebular Abundances}

\subsection{Line Ratios and Diagnostic Diagrams}
Analysis of the \hii region line fluxes followed the same
procedures as described in \citet{vHS97a} and \citet{vSHOB98}.
Briefly, emission line strengths were measured in the 1-d 
spectra and then corrected for underlying Balmer absorption
and for reddening.  The intrinsic Balmer line strengths were
interpolated from the tabulated values of \citet{HS87} for case B
recombination, assuming N$_e$ = 100 cm$^{-3}$ and T$_e$ = T$_{[OIII]}$ 
(see Section 3.2).  Assuming a value of $R = A_V/E_{B-V} = 3.1$, the galactic 
reddening law of \citet{S79} as parameterized by \citet{H83} was adopted to
derive the reddening function, $f(\lambda)$, normalized at H$\beta$.
For those \hii regions with detected [O III] $\lambda$4363, the
temperature was then recalculated from the corrected line strengths
and a new reddening coefficient was produced.
 An underlying Balmer absorption with an equivalent
width of 2 \AA~was assumed in the few instances where the reddening coefficient
was significantly different when derived from the observed line ratios of
H$\alpha$/H$\beta$ and H$\gamma$/H$\beta$.

The reddening corrected line intensities relative 
to H$\beta$ and the reddening coefficients, c$_{H\beta}$, 
for each \hii region are listed in Table \ref{tab:flux}.
In this and subsequent tables, each H II region is identified by 
its east--west and north--south offsets from the galaxy center 
(in arcsec, north and east are positive) and by its slit
number (Table \ref{tab:slits}). 
 The error associated with each relative line intensity
was determined by taking into account the Poisson noise in the line, the
error associated with the sensitivity function, the contributions of
the Poisson noise in the continuum, read noise, sky noise, and flat
fielding or flux  calibration errors, the error in setting the
continuum level (assumed to be 10\% of the continuum level),
and the error in the reddening coefficient.  

\begin{figure}
\plotone{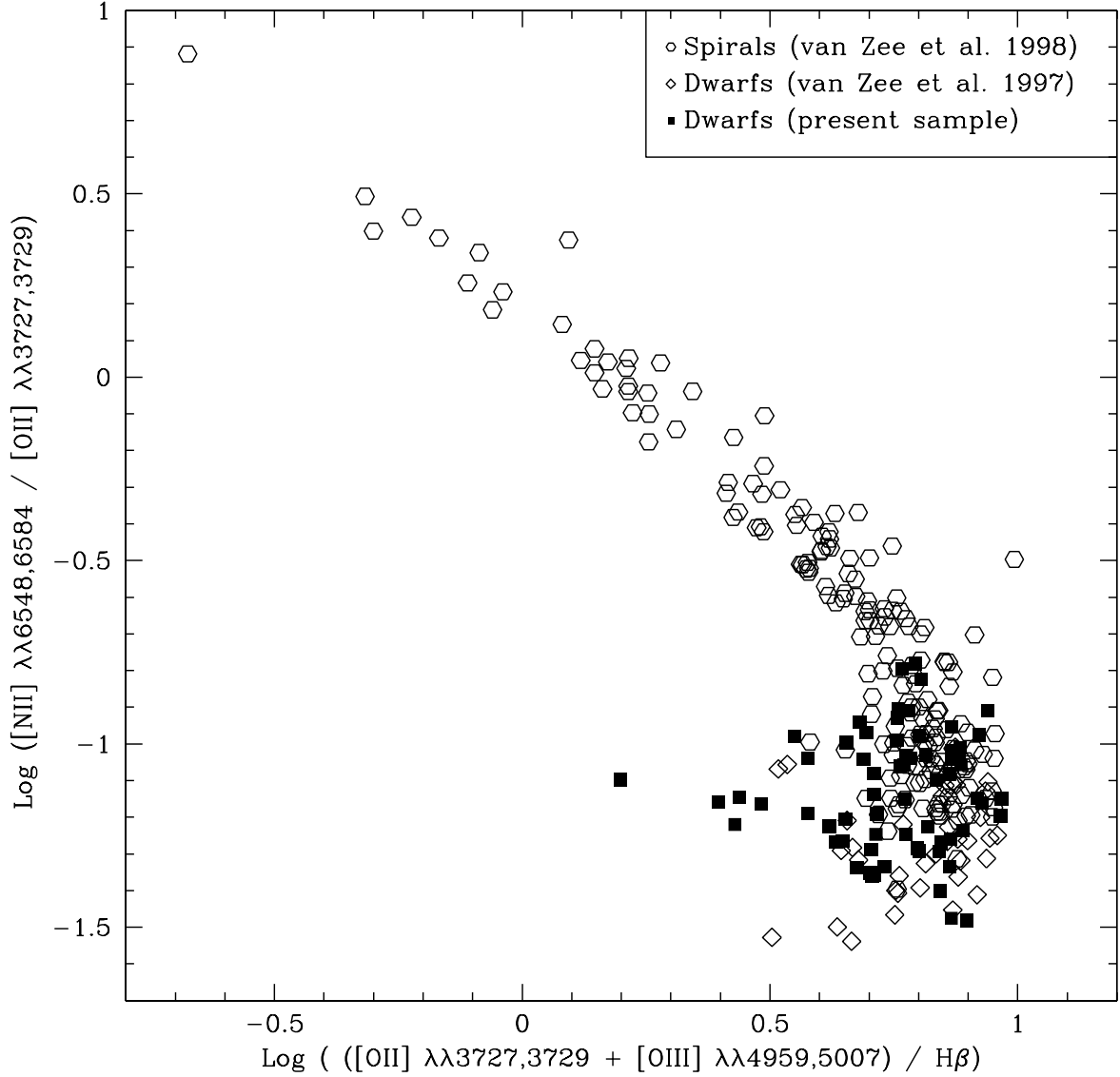}
\caption[]{
Diagnostic diagram for nitrogen and oxygen line ratios in dwarf irregular
(present sample; van Zee et al.\ 1997) and spiral galaxies (van Zee et al.\ 1998).  
As expected, the low abundance HII regions in dwarf irregular galaxies
have low [N II]/[O II].  However, the high metallicity H II regions in 
spiral galaxies trace a fairly narrow locus in this diagram while
the low abundance H II regions in dwarf irregular galaxies exhibit a 
wide range of [N II]/[O II] for a given value of R$_{\rm 23}$. 
\label{fig:diag_no} }
\end{figure}

All of the \hii regions in this sample fall within the normal
locus of line ratio values in standard diagnostic diagrams.  For example,
the ratio of [N II]/[O II] is shown as a function of the
strength of the strong oxygen lines (R$_{23} \equiv$ ([O II]+[O III])/H$\beta$) in Figure
\ref{fig:diag_no}.  All of the \hii regions in this sample
have relatively weak [N II] emission, as expected for
low metallicity \hii regions.   Also included in Figure \ref{fig:diag_no}
are spiral galaxy \hii regions \citep{vSHOB98} and \hii regions
in low surface brightness dwarf galaxies \citep{vHS97a} which were
observed and analyzed in a similar manner to those in the present study.
The high metallicity \hii regions trace a narrow locus in this
diagnostic diagram while the dwarf galaxy \hii regions exhibit a
wide range of [N II]/[O II] for a given value of R$_{\rm 23}$. 
Such a range is expected for H II regions with a wide range of 
ionization parameters \citep[e.g.,][and see Section \ref{sec:strongline}]{M94}.

\subsection{Temperature and Density Determinations}

When possible, the electron temperature of the ionized gas is
derived from the reddening corrected line strengths of 
the [O III]  $\lambda$5007, $\lambda$4959, and $\lambda$4363 lines 
(Table \ref{tab:ratio}).  The procedure to determine the electron 
temperature of the O$^{++}$ zone is well described in \citet{O89} and 
will not be repeated here. A version of the FIVEL program of 
\citet{DDH87} was used to compute T$_e$ from the [O III] line ratios.
In addition, since numerical models of \hii regions have shown that there
are significant differences between the temperatures in the
high- and low-ionization zones \citep{S80}, we adopt the
approach taken by \citet{Pe92}, who use the \hii region
models of \citet{S90} to derive an approximation of the electron 
temperature in the O$^{+}$ zone:
\begin{equation}
T_e(O^+) = 2 [ T_e^{-1}(O^{++}) + 0.8]^{-1},
\end{equation}
where $T_e$ is the electron temperature in units of $10^4$ K.

In many cases, [O III] $\lambda$4363 either is not detected or is
 contaminated by the nearby Hg $\lambda$4358 night sky line.  In
these cases, we use ionization models to approximate the oxygen
abundance and the electron temperature.
Because oxygen is one of the dominant coolants in low metallicity
\hii regions, the electron temperature
depends strongly on the oxygen abundance.  Thus, we determine
the electron temperature necessary to obtain the oxygen
abundance as determined by strong-line methods (see Section \ref{sec:strongline})
and adopt that value for subsequent analysis of other
atomic abundance ratios.
While the errors in the electron temperatures derived by this method
are quite large, relative atomic abundances (such as N/O) are less
sensitive to choice of electron temperature \citep[e.g.,][]{KS96}.
We caution, however, that systematic errors in the strong-line abundance
calibration will introduce systematic errors in these temperature
estimates as well.  For H II regions where the
electron temperature was derived from the assumed oxygen abundance,
the error in the electron temperature was set to yield the appropriate error
(as determined by the calibration process) in the computed oxygen abundance.
The electron temperatures for the O$^{++}$ zone adopted for the 
subsequent abundance analysis are tabulated in Table \ref{tab:abund}.

The density sensitive line ratio [S II] $\lambda6717/6731$ is tabulated
for each H II region in Table \ref{tab:ratio}.
The majority of the H II regions in this sample are within the low density limit
(I($\lambda$6717)/I($\lambda$6731) $>$ 1.35).  In calculating the
emissivities for the abundance analysis, an electron density of 100 cm$^{-3}$ was assumed 
unless the [S II] line ratios were below the low density limit, in which
case the FIVEL program \citep{DDH87} was first used to calculate the
electron density from the [S II] line ratios.

\subsection{Direct Oxygen Abundances}
\label{sec:direct}
For \hii regions with a solid detection of [O III] $\lambda$4363,
the oxygen abundances tabulated in Table \ref{tab:abund} are 
derived from the observed line strengths (Table \ref{tab:flux}) and  
the relevant emissivity coefficients (based on the electron temperature 
(Table \ref{tab:abund}) and electron density) as calculated by the 
FIVEL program \citep{DDH87}.  Even in these relatively high signal-to-noise 
ratio spectra, the errors in the derived abundances are dominated by 
the errors inherent in the electron temperature derivation.
Fourteen of the 21 galaxies in this sample have at least one \hii region
with a solid $\lambda$4363 detection (25 \hii regions in total).
As found in other low mass galaxies \citep[e.g.,][]{KS97}, oxygen
abundances derived from multiple \hii regions within the same
galaxies are quite similar (see Section \ref{sec:global} for further discussion).
As the abundances calculated from direct detection of the temperature
sensitive lines are less likely to be affected by significant
systematic errors, these abundances are given  more
weight in the mean abundances tabulated in Table \ref{tab:gals}.

\subsection{Strong-Line Oxygen Abundances}
\label{sec:strongline}
\subsubsection{Photoionization Model Grids}
The majority of spectra in this sample are of \hii regions with 
low ionization parameters; thus, despite high signal--to--noise
ratios in the strong lines, the [O III] $\lambda$4363 line is often
too weak to use to calculate the electron temperature directly. 
 However, the behavior of the strong optical 
oxygen line intensities ([O II] $\lambda3727$ and [O III] $\lambda\lambda 4959,5007$)
as a function of oxygen abundance has been examined extensively via semi--empirical
methods \citep[e.g.][]{S71,PEBCS79,ACJV79,EP84,DE86,S89}
and derived from H II region ionization models 
\citep[e.g.,][]{M91,O97,KD02}.  At low metallicities,
the dominant coolant is the collisionally excited Lyman series, 
and thus the total oxygen line intensity (R$_{23}$)
increases as the abundance increases.  As the metallicity increases, however, the 
infrared fine structure lines, such as the 52$\mu$ and 88$\mu$ oxygen lines and 
the C$^+$ 158$\mu$ transition, begin to dominate the cooling.  At an oxygen 
abundance of approximately 1/3 solar, the intensity of the optical oxygen 
lines reaches a maximum and then declines as the oxygen abundance increases.  
Thus, the relationship between R$_{\rm 23}$ and oxygen abundance is double valued.
As described by \citet{ACJV79}, this degeneracy can be broken with additional 
information, such as the relative strength of the nitrogen and oxygen lines.
The low [N II]/[O II] line ratios (Table \ref{tab:ratio}; Figure \ref{fig:diag_no}) 
indicate that all of the \hii regions in the present sample are located on the
lower branch of the R$_{\rm 23}$ relation.

The lower branch of the theoretical model grid of 
\citet{M91} is shown in Figure \ref{fig:r23}.  
In this and other similar photoionization model calibration methods \citep[e.g.,][]{KD02},
the geometry of the H II region (represented here by the average 
ionization parameter, \=U, the ratio of ionizing photon density 
to particle density)  introduces a further spread in the estimated abundance 
for a given R$_{23}$.  Thus, the process of determining an empirical
abundance from the strong line ratios requires knowledge of both the
total oxygen line intensity and the ratio of [O III] to [O II] to
estimate the ionization parameter.  Oxygen abundances derived from
the photoionization model grid of \citet{M91} are tabulated for
all of the \hii regions in Table \ref{tab:abund}.

Most strong-line abundance calibrations are derived from photoionization
models of zero-age \hii regions \citep[e.g.][]{M91,KD02}.  Thus, systematic abundance 
errors may be introduced when \hii regions age, as both the ionization parameter 
and shape of the ionizing spectrum will evolve \citep[e.g.,][]{SL96,O97}. 
As discussed in \citet{vSH05}, such effects may
be significant for \hii regions with log ([O III]/[O II]) $<$ -0.4.
In this sample, UGC 1175 -004+007, UGC 2023 -028+023, UGCA 292 +026+004, 
and UGC 9992 +007+010 all lie within this region of high uncertainty; 
however, due to the lack of detailed age information for these particular 
\hii regions, we do not include age effects in the present analysis.
Fortuitously, several \hii regions were observed in all four of these galaxies,
and thus the tabulated mean abundances (Table \ref{tab:gals}) do
not rely heavily on the results from these four extremely low 
ionization parameter \hii regions.

\subsubsection{Comparison of Direct and Strong Line Abundance Calibrations}

\begin{figure}
\plotone{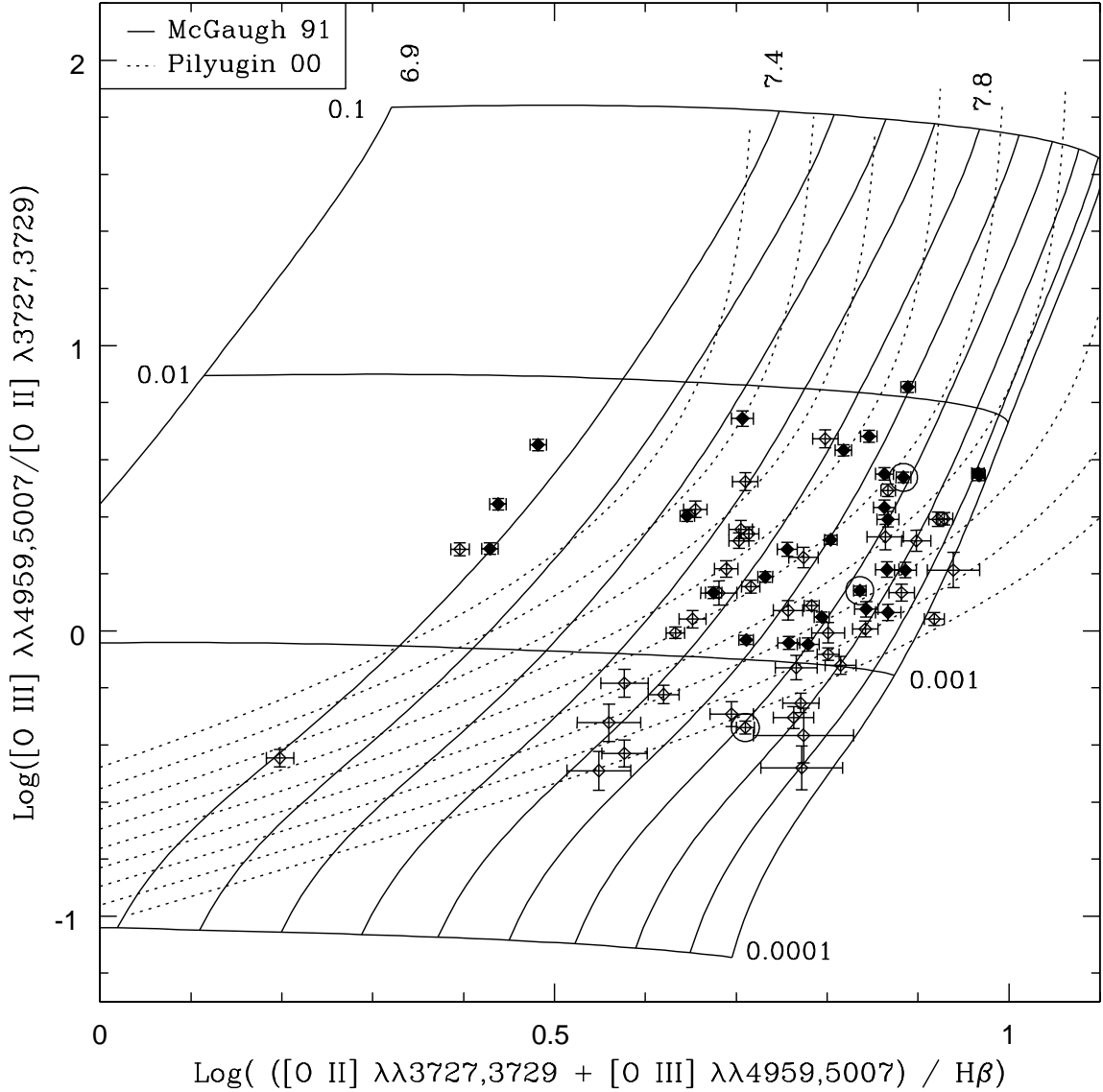}
\caption[]{
Model grid of the lower branch of the R$_{\rm 23}$ relation from
McGaugh (1991) (solid lines).  Also shown are the empirical 
metallicity calibration relations derived by Pilyugin (2000) 
for 7.4 $<$ 12 + log(O/H) $<$ 8.2 (dashed lines).  
The locations of H II regions in this sample are marked;
filled diamonds represent H II regions with solid detections
of [O III] $\lambda$4363.  The circled points correspond 
to the representative H II regions shown in Figure 1.
The five extremely low metallicity points 
(12 + log(O/H) $\sim$ 7.35) are all observations 
of H II regions in UGCA 292.  
\label{fig:r23} }
\end{figure}

As discussed in \citet{vSH05}, the \citet{M91} photoionization
models appear to trace out the appropriate shape for iso-metallicity contours in
the abundance diagnostic diagram, but there is evidence of a systematic
offset between direct and empirical calibration methods.  For example,
in Figure \ref{fig:r23}, the five \hii regions in UGCA 292 form a tight
locus around the 12+log(O/H) = 7.35 iso-metallicity contour while direct
abundance calculations for 3 of these \hii regions yield 
12+log(O/H) $\sim$ 7.32 $\pm$ 0.03. 

\begin{figure}
\epsscale{0.9}
\plotone{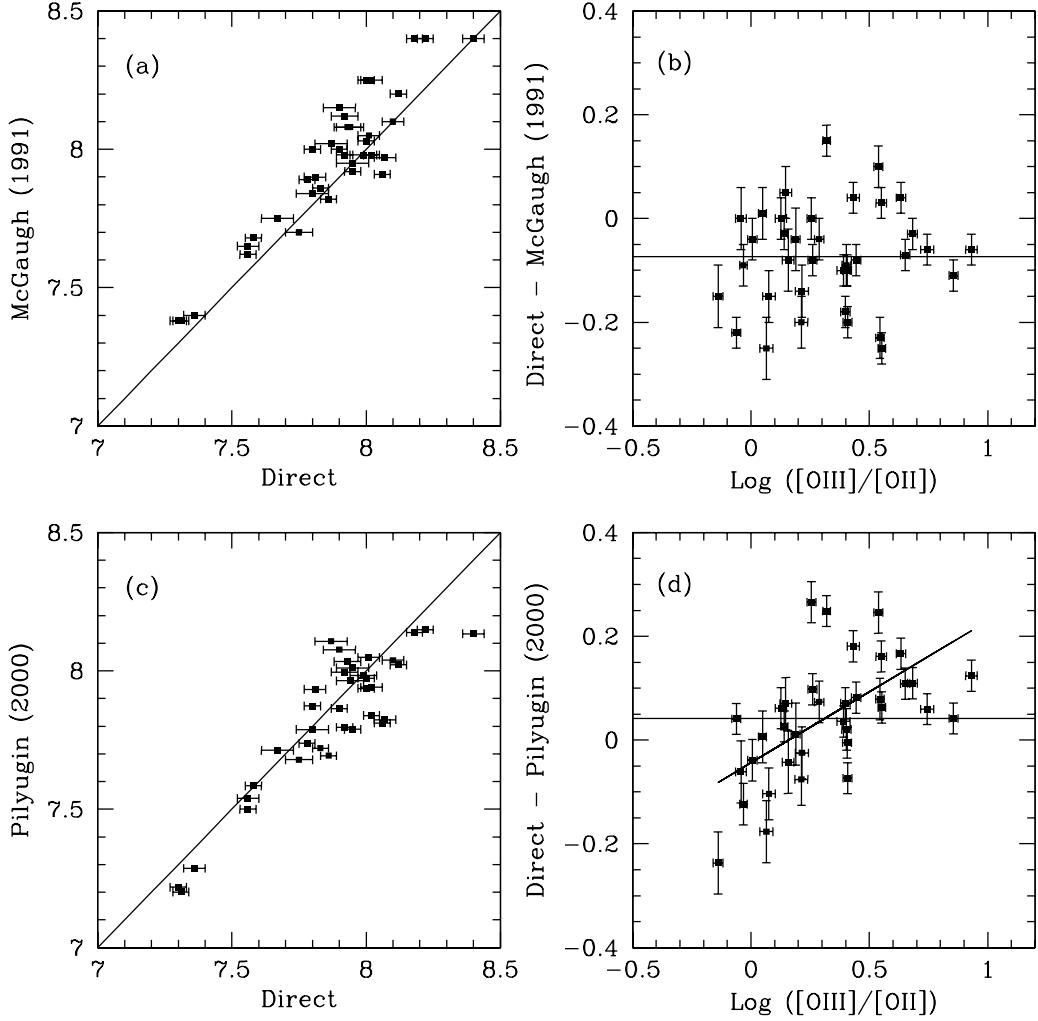}
\caption[]{
Comparison of empirical and direct abundance calibrations for H II regions
in dwarf irregular galaxies with detected [O III] $\lambda$4363 emission.  
The plots include H II regions in the present sample as well as observations
reported in van Zee et al.\ (1997).  Both the McGaugh (1991) and Pilyugin (2000)
calibrations have systematic offsets from the direct oxygen abundance
calculation.  (top panels) The photoionization models from McGaugh (1991) 
appear to overpredict the oxygen abundance by 0.07 dex.  (bottom panels) 
The p-method (Pilyugin 2000) appears to underpredict the oxygen abundance 
by $\sim$0.04 dex, but there is a systematic correlation with the ionization 
parameter in the sense that the p-method significantly (0.1 - 0.2 dex) overpredicts 
the oxygen abundance for low excitation H II regions.
\label{fig:comp}
}
\end{figure}

Figure \ref{fig:comp} compares the direct abundance determinations
(Section \ref{sec:direct}) with the strong line abundance determinations
(Section \ref{sec:strongline}.1) for the 25 \hii regions in the
present sample with high signal-to-noise ratio detections of
[O III] $\lambda$4363 as well as 12 \hii regions from
\citet{vHS97a} which were analyzed in a similar manner.
The statistical offset between direct abundance calculations
and the \citet{M91} model grid is -0.07 $\pm$ 0.10 dex for these 37 \hii regions.  
If the high ionization parameter \hii regions of \citet{ITL94}, \citet{TIL95},
\citet{ITL97}, and \citet{IT98}
are also included, the statistical offset is -0.08 $\pm$ 0.11 dex.
Thus, the \citet{M91} strong line abundance calibration appears to 
overpredict the oxygen abundance relative to direct abundance calculations.
However, the statistical offset between direct abundances and \citet{M91}
calibration does not appear to have systematic trends in
regards to signal-to-noise ratio, oxygen abundance, or
ionization parameter.  Thus, to place the strong line abundances
and direct abundances on the same approximate calibration scale,
it may be appropriate to subtract 0.07 dex from the tabulated
values.

Offsets between photoionization model grids and direct calibration
methods have been noted by many authors;
recently, \citet{P00} proposed empirical calibrations based on 
\hii regions where [O III] $\lambda$4363 has been 
detected.  However, such purely empirical 
calibrations potentially suffer from severe selection effects since the
majority of high quality abundances available in the literature
are biased toward \hii regions with high ionization parameters,
since these \hii regions will have stronger [O III] line
intensities for a given abundance.  For example, 
\hii regions in blue compact dwarf galaxies typically have 
ionization parameters  between 0.01 and 0.1 \citep[e.g.,][]{ITL97,IT98}.
The present sample significantly extends the number
of \hii regions of moderate ionization parameters and
direct abundance determinations.   Similar to the \hii
regions in \citet{vHS97a}, the \hii regions in the present
sample tend to have low average ionization parameters, \={U} $\sim$ 0.001 -- 0.01; 
only one \hii region in this sample, CGCG 007-025 -004+000,
has an ionization parameter comparable to \hii regions
in \citet{ITL97,IT98}.

The model grid for the \citet{P00}
empirical calibration based only on high ionization
parameter \hii regions (the p-method) is shown in Figure \ref{fig:r23}.  
In the high ionization parameter
region, there is an offset of approximately 0.1 dex between
the \citet{M91} and \citet{P00} calibrations, as expected
since the \citet{M91} photoionization model grid appears
to overestimate the oxygen abundance by approximately this
amount.  However, in the lower ionization parameter regions
of this diagnostic diagram, the \citet{P00} empirical
calibration  deviates significantly from the photoionization
grid, in a non-physical manner.  Unfortunately, this is
precisely the region of the diagnostic diagram where one
is most likely to need to use a strong line abundance
indicator, as the [O III] lines are weaker and thus the
temperature sensitive lines will have lower signal--to--noise ratios.  

A comparison of direct abundance calibrations and those derived
by the p-method for the 37 moderate
ionization parameter \hii regions with solid [O III] $\lambda$4363
detections (present sample; van Zee et al.\ 1997) is shown in Figure \ref{fig:comp}c.  
The p-method yields an oxygen abundance
which is statistically 0.04 $\pm$ 0.11 dex lower than
the direct abundance calculations in this sample.  If the
high ionization parameter blue compact dwarf galaxies of Izotov \& Thuan are
included, the statistical deviation of the p-method is
0.01 $\pm$ 0.14 dex (a lower statistical offset is expected
when the blue compact dwarf galaxies are included since 
these same \hii regions were used to derive the empirical calibration).
However, as can be seen in Figure \ref{fig:comp}d,
the deviation from the direct abundance calculations is 
strongly correlated with the ionization parameter.
Thus, \hii regions with high ionization parameters 
will be systematically underestimated by the p-method
while \hii regions with low ionization parameters (i.e.,
the most likely type of \hii region for which 
a strong-line calibration will be needed) will
be systematically overestimated by the p-method.
Although the values are tabulated in Table \ref{tab:abund},
we elect not to use the p-method abundances for the \hii regions
in this sample due to these systematic calibration errors.

\subsection{Relative Abundances of Nitrogen, Neon, Sulfur, and Argon}

\subsubsection{Ionization Correction Factors}
For all atoms other than oxygen,  derivation of atomic
abundances requires the use of ionization correction factors (ICFs) to account
for the fraction of each atomic species which is in an unobserved ionization state.
To estimate the nitrogen abundance, we have
assumed that N/O = N$^+$/O$^+$ \citep{PC69}.
 To estimate the neon abundance, we have assumed that
Ne/O =  Ne$^{++}$/O$^{++}$ \citep{PC69}.
To determine the sulfur and argon abundances, we are forced to adopt an
ICF from published H II region models to correct for the unobserved S$^{+3}$
and Ar$^{+3}$ states.   We have adopted analytical forms of the ICFs as given by
\citet{TIL95}.
Given the large uncertainties in the ICF,
we do not report a sulfur abundance for H II regions where S$^{++}$  $\lambda 6312$
is not detected.  

\subsubsection{Abundance Ratios}

\begin{figure}
\epsscale{1.0}
\plotone{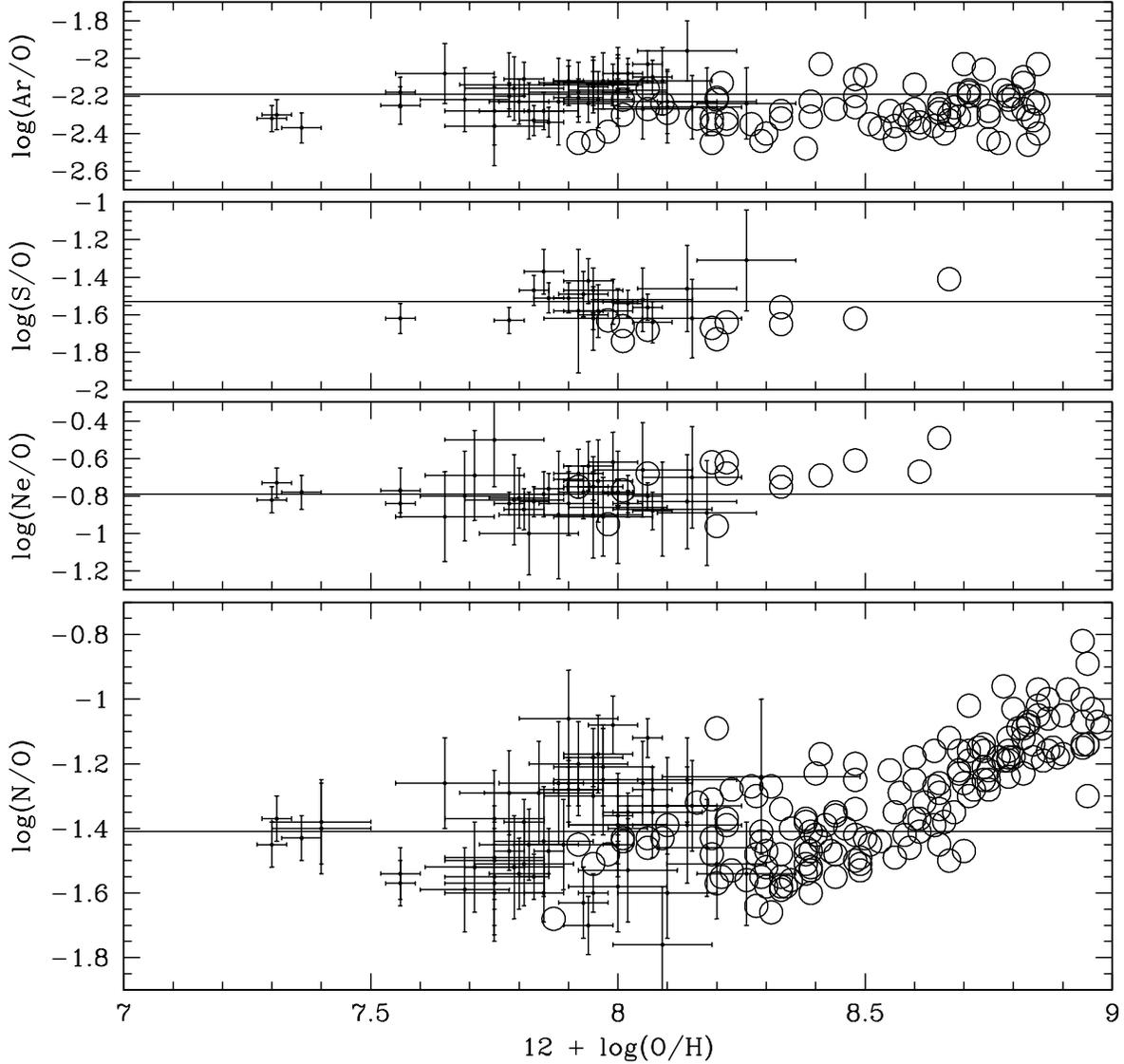}
\caption[]{
H II region elemental abundances for nitrogen, neon, sulfur, and argon
as a function of oxygen abundance.  The figures include H II regions 
of dwarf irregular galaxies from the present sample as well as 
observations reported in van Zee et al.\ (1997); open circles denote
H II regions in spiral galaxies (van Zee et al.\ 1998).  As expected,
the alpha elements (neon, sulfur, and argon) show no trend with
increasing oxygen abundance, while the N/O ratio is approximately
constant at low abundances (primary nitrogen), but then increases 
as the oxygen abundance increases (secondary nitrogen).
\label{fig:abund}
}
\end{figure}

The abundance ratios of nitrogen, neon, sulfur, and argon relative to 
oxygen are tabulated in Table \ref{tab:abund} and are shown graphically
in Figure \ref{fig:abund}.  Also include in these figures are
additional dwarf irregular galaxy H II regions \citep{vHS97a}
and HII regions in spiral galaxies \citep{vSHOB98}, all of which
were analyzed in a similar manner.   The \hii regions in this
sample range from those with extremely low metallicity 
(e.g., UGCA 292, 12 + log(O/H) = 7.32) to those that are a 
modest fraction of solar (e.g., UGC 3647, 12 + log(O/H) = 8.07).

As expected, the relative abundance ratios of the alpha elements
(neon, sulfur, and argon) to oxygen is constant as a function
of oxygen abundance.  For this sample, the mean log(Ne/O)
abundance ratio is -0.79 $\pm$ 0.09, the mean log(S/O) is
-1.53 $\pm$ 0.09, and the mean log(Ar/O) is -2.19 $\pm$ 0.09.
These values are comparable to the values reported for
starbursting dwarf galaxies \citep[e.g.,][]{TIL95,ITL97,IT98}
and for other dwarf irregular galaxies \citep[e.g.,][]{vHS97a}.

At moderate and high metallicity, nitrogen is a secondary
element \citep[log(N/O) increases linearly with log(O/H), e.g.,][]{TPF89,VE93,HEK00};
however,  at low metallicity, nitrogen behaves like a primary
element with an approximately constant log(N/O) ratio
for 12 + log(O/H) $<$ 8.3 \citep[e.g.,][]{VE93,vSH98,HEK00}.  However, while the mean
value is approximately constant at low metallicity,
the range of observed log(N/O) increases significantly
at 12 + log(O/H) $>$ 7.9 \citep[see also][]{KS96,KS98}.  
In this sample of \hii regions, the mean log(N/O) ratio is -1.41 $\pm$ 0.15, 
which is slightly higher than that found in extremely 
low metallicity \hii regions \citep[e.g.,][]{TIL95,ITL97,IT98},
but comparable to the observed values for several nearby
low metallicity dIs \citep[e.g.,][]{SCM03,VP03}.

\subsection{Global Abundances}
\label{sec:global}
\begin{figure}
\includegraphics[angle=-90,width=17cm]{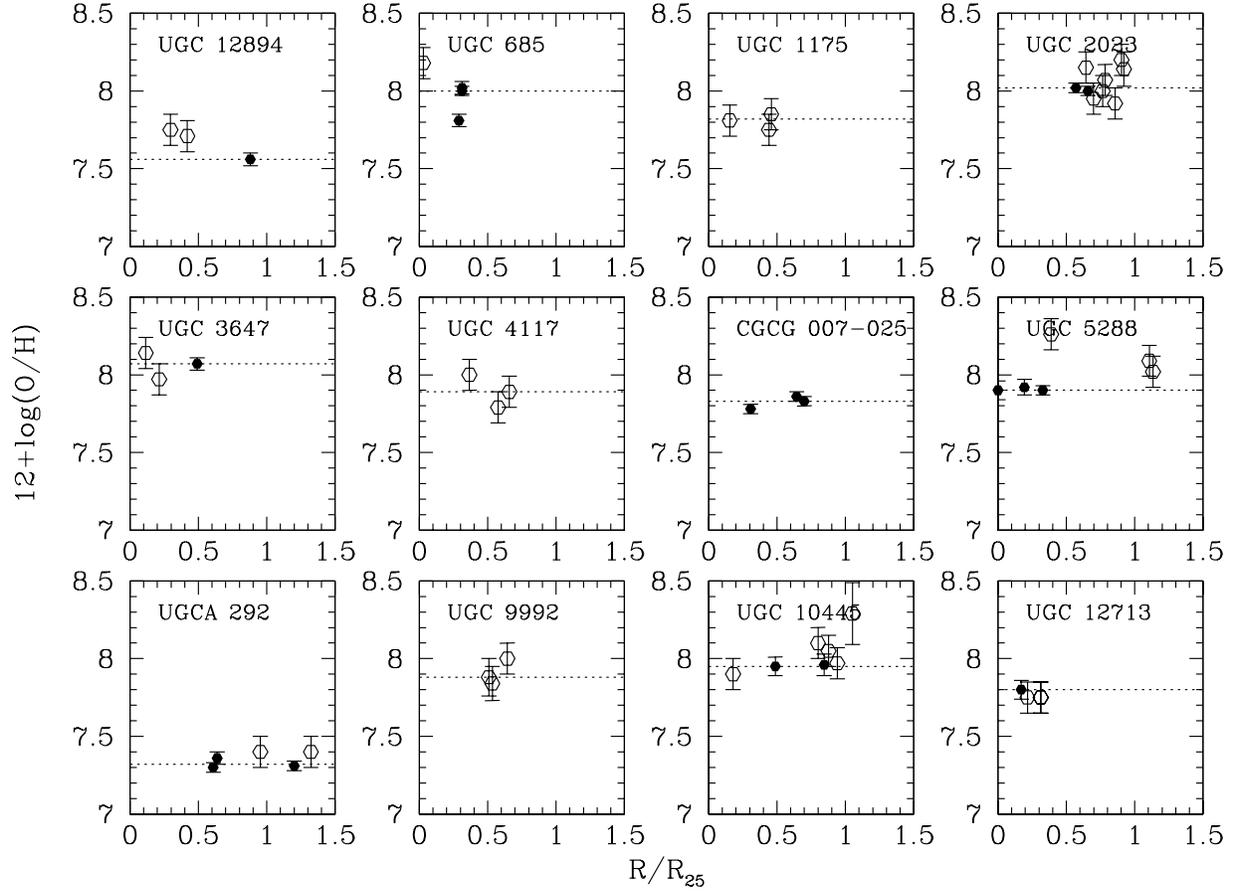}
\caption[]{
H II region abundances as a function of (normalized) galactic
radius; filled symbols represent H II regions with solid
detections of [O III] $\lambda$4363.  
In  most instances, the \hii region abundances 
are similar at all radii.  Note that the apparent trend
in UGC 12894 may be a result of an overestimate of the
inner H II region oxygen abundances (strong line method).
The dashed line indicates the mean metallicity listed in Table 1.
\label{fig:grad}
}
\end{figure}

Multiple \hii regions were observed in 16 of the 21 
galaxies. Figure \ref{fig:grad} shows the observed
oxygen abundances as a function of (normalized)
galactic radius for the 12 galaxies with three
or more observations which sample a range of
galactic radii.
In most instances, the derived elemental 
abundances are very similar (within the formal errors) 
within each galaxy.  The one exception may be UGC 12894,
but this plot may be misleading since
the abundances of the inner \hii regions are derived by the 
strong line method, and thus may be systematically too high 
(see Section \ref{sec:strongline}).
Thus, in most of these dwarf irregular galaxies
both oxygen and nitrogen appear to be well mixed.
Similar results have been found in other extensive studies of 
dwarf irregular galaxy abundances \citep[e.g.,][]{KS96,KS97,VP98,LS04}.
Given their small dimensions and spatially homogeneous star formation 
histories \citep[e.g.,][]{vZ01}, such uniform abundances are perhaps 
expected in dwarf irregular galaxies. Theoretical
models of dispersal and mixing also suggest that 
the mixing time scales should be short, and that dwarf
galaxies should be well mixed \citep[e.g.,][]{RK95,TT96}.
Global oxygen and nitrogen-to-oxygen ratios are
summarized in Table \ref{tab:gals}.  In all instances,
abundances derived from direct temperature measurements
are given higher weight in the global averages.

Seven of the 21 galaxies in this sample have oxygen 
and nitrogen abundances previously reported.  In general, the abundance
measurements agree well for those galaxies with
high signal--to--noise ratio observations both here and
in the literature [UGC 4483, \citet{STKGT94},
CGCG 007-025, \citet{KPGLP04,IT04}, and UGCA 292,
\citep{vZ00a}].  However, the present observations indicate
a slightly lower oxygen abundance for UGC 9240
\citep[7.95 $\pm$ 0.03 as compared to 8.01 $\pm$ 0.03,][]{HGO02}
despite both observations including detection of
the temperature sensitive [O III] $\lambda$4363 line.
Finally, as expected for comparisons of direct and
strong line abundances, the present observations
yield significantly lower oxygen abundances for
both UGC 2023 and UGC 3647 \citep{HH99}.
It is important to note that 
despite the differences in absolute abundance
calibration, the observations reported in \citet{HH99} also 
indicate that the oxygen abundances are spatially homogeneous 
in these two galaxies.

\section{Chemical Enrichment of Gas-Rich Dwarf Irregular Galaxies}

\subsection{The Baseline Model: Closed Box Chemical Evolution}

In the following discussion, 
the simple model of closed box chemical evolution 
will be used as the baseline for comparison of
the observed chemical enrichment and theoretical
models.  As described in detail by \cite{E90},
a simple closed box model sets the limit between
the permitted and forbidden zones in abundance
diagnostic diagrams.  The simple closed box model
assumes both instantaneous recycling and that the
products of stellar nucleosynthesis are neither
diluted by infalling pristine gas nor lost to the
system via outflow of enriched gas \citep[e.g.,][]{TA71,SS72}.  
In a closed box model, the expected elemental abundance is
related only to the yield (p) and the gas
mass fraction (f):
\begin{equation}
z = p~{\rm ln}~(1/f)
\end{equation}
where the gas mass fraction is the ratio of
the gas mass to total baryonic mass of the
system:
\begin{equation}
f = g/(g+s)
\end{equation}
where $g$ is the mass in gas and $s$ is the mass
in stars.  The elemental yield is estimated from
models of stellar nucleosynthesis and an initial
mass function (IMF).  Recent stellar evolution models
indicate that for some elements, such as oxygen and
nitrogen, the yield may be metallicity dependent and 
may depend on the stellar kinematics as well \citep[e.g., rotation,][]{MM02}.
In the following discussion, we adopt a \citet{S55} IMF
with an upper and lower limit of 120 and 2 M$_{\odot}$,
respectively, and the integrated yields for
stars with rotation from \citet{MM02}: 
p$_{\rm O}$ = 7.4 $\times$ 10$^{-3}$ and p$_{\rm N}$ = 1.1 $\times$ 10$^{-4}$.  
Changes in the upper mass limit of the IMF will primarily affect the
yield of oxygen while changes in the lower mass limit
of the IMF will primarily affect the yield of
nitrogen.  For the abundance range of interest (12 + log(O/H) $<$ 8.4),
neither the oxygen nor the nitrogen yields are
sensitive to metallicity for this choice of IMF
\citep[most modern stellar nucleosynthesis models now include 
primary nitrogen production at low metallicities, e.g.,][]{TWW95,WW95,MM02}.
However, while the ratio of N/O is only moderately
affected, both oxygen and nitrogen yields are sensitive 
to the stellar kinematics; the stellar nucleosynthesis models 
with stellar rotation adopted here have higher yields than 
models without rotation.

A closed box enrichment model requires knowledge of both 
the gas mass and stellar mass.  In this calculation,
the gas mass includes the atomic component
with a correction for neutral helium
[M$_{\rm gas}$ = 1.3 $\times$ M$_{\rm HI}$], 
but does not include molecular gas since it is
difficult to detect CO in low mass galaxies  
\citep[e.g.,][]{EEM80,TKS98}, and, further, the correction from CO
to H$_2$ column density is highly uncertain for low metallicity galaxies
\citep[e.g.,][]{VH95,W95}. 
The stellar mass is derived from the
B-band luminosity and the mass-to--light ratio
calibration of \citet{vZ01}:
\begin{equation}
{\rm log} \Gamma_b = 2.84(B - V) - 1.26
\end{equation}
which was derived from the \citet{BC96} code for a quasi-continuous
star formation rate model.  We adopt the typical mass--to--light
ratio of 0.75 (B-V $\sim$ 0.4) for the handful of galaxies with no
optical colors available. 

\begin{figure}
\epsscale{0.9}
\plotone{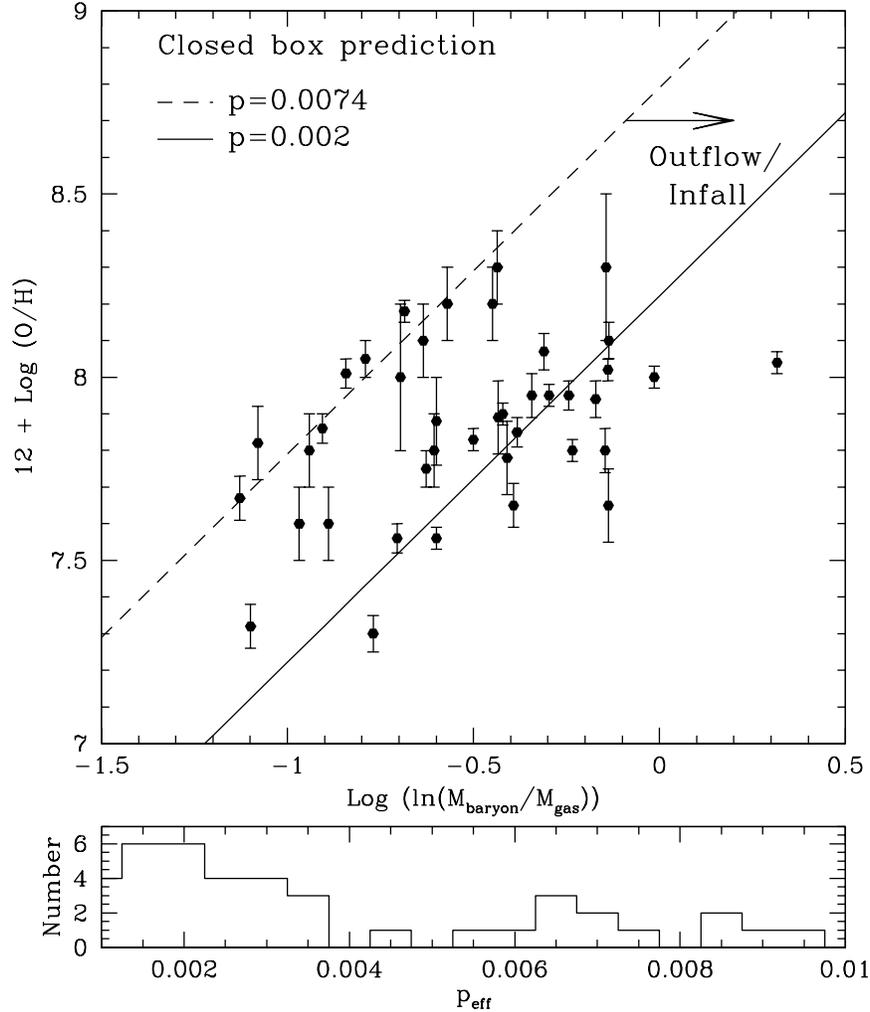}
\caption[]{
Comparison of the observed oxygen abundance and that predicted
by simple closed box chemical evolution models with instantaneous
recycling and constant star formation rates.  The dashed line
indicates the expected one-to-one trend if dwarf irregular
galaxies are closed boxes with an oxygen yield appropriate
for low metallicity galaxies with a Salpeter IMF (p = 0.0074, 
Meynet \& Maeder 2002).  The solid line indicates the expected 
one-to-one trend if dwarf irregular galaxies evolve as open boxes
with an effective yield $\sim$1/4 of the true yield.
The bottom panel shows the histogram of effective yields
for this sample of dwarf galaxies.  While the majority 
of dIs appear to evolve as open boxes (due to either
infall of pristine gas or outflow of enriched material),
several of the dwarf irregular galaxies in this sample appear to be 
consistent with closed box chemical evolution.  
\label{fig:yield}
}
\end{figure}

A comparison of the observed oxygen abundances to
those predicted by closed box models is shown in 
Figure \ref{fig:yield}.
In this and subsequent figures, the galaxy sample includes both the present
study and 19 galaxies from \citet{vHS97a,vSH05} with distances
converted to H$_0$=75 km s$^{-1}$ Mpc$^{-1}$ for
consistency (Table \ref{tab:others}). 
 While a large number of oxygen abundances
now exist in the literature \citep[see, e.g., compilation
in][]{vSH05}, at this time we restrict our analysis to 
this large uniform sample for which complementary deep optical
and H I imaging exists.  

Figure \ref{fig:yield} indicates that several of the galaxies in 
this expanded sample are consistent with closed box chemical
evolution \citep[as also noted in][]{vHS97b}.
However, the majority of galaxies in this sample appear
to have oxygen abundances less than that expected by
closed box models.  The effective yield, p$_{\rm eff}$,
is defined as the yield that would be deduced if the
galaxy was assumed to be a simple closed box.
A histogram of the effective yields for this sample
is shown in the bottom panel of Figure \ref{fig:yield}.  
The majority of galaxies appear to have an effective
yield $\sim$1/4 of the true yield (p$_{\rm eff}$ $\sim$ 0.002). 

While it is possible to minimize the offset between the
closed box prediction and the observed abundances by
tweaking the input model parameters, such 
efforts require either unusually high molecular--to--atomic gas
ratios or unusually low stellar mass--to--light ratios.
For instance, the majority of the galaxies can be brought
into agreement with closed box chemical evolution models
if we assume that the (unknown) molecular gas component is $\sim$3
times the atomic component.  Alternatively, the majority of the 
observations agree with a closed box estimate if the typical stellar 
mass--to--light ratio is 0.2.  An intermediate result of
M$_{H_2}$ $\sim$ M$_{HI}$ and $\Gamma_B$ $\sim$ 0.4
will also force the closed box model to fit the observations
for the majority of the galaxies.  However, even with adhoc 
adjustments of the input parameters, it is not possible to
fit simultaneously all of the galaxies to a closed box model
since those that previously appeared to be `closed' are now 
overenriched relative to the model.  Furthermore, none of these
adhoc adjustments are motivated by physical processes.
Thus, it appears that inflow of pristine gas or outflow 
of enriched gas may be the most straightforward explanation
for the range of observed abundances in many low mass dwarf galaxies.

\subsection{Effects of Gas Flows and Time Delays on Chemical Enrichment}

An excellent overview of the effects of gas flows on 
observed chemical abundances is provided by \citet{E90}.
In brief, both infall and outflow of well mixed 
material will result in effective yields that
are less than the true yields as the enriched material is
either diluted (infall) or lost from the system (outflow).  
In this section, we discuss physical processes 
that can substantially affect chemical enrichment and summarize
expected effects on the observed oxygen and nitrogen abundances.
For the purposes of the following discussion, nitrogen
will be considered a primary element, as appears to be 
appropriate at low metallicity \citep[e.g.,][]{EP78}.

{\it (i) Inflow of pristine gas.}  If material added to
the system is pristine, the effect of gas infall is to dilute 
the enriched gas and thus reduce the effective yield.
Infall of pristine gas should have no effect on relative abundance
ratios, such as N/O, since dilution is assumed to
occur instantly and uniformly and both the oxygen and
nitrogen abundances will be lower than predicted by
closed box models.  However, if a system undergoes
significant infall, differences between closed box 
abundances and observed abundances may be related to 
the gas mass fraction since galaxies with `excess' gas 
may be more likely to have had significant infall.

{\it (ii) Infall of enriched gas.}  If the new material
added to the system is pre-enriched, the effect of infall depends
on whether the metallicity is higher or lower than that 
of the system itself \citep[e.g.,][]{E90}.
If the new material has lower metallicity, the
effective yield will be less than the true yield.
In addition, if the new material has similar abundance ratios
as the existing gas (as is expected for primary elements, 
for example), the effect on relative abundance ratios,
such as N/O, is minimal since the new and old material
are similarly enriched.  The effects are more complicated
if the new material consists of both primary and secondary nitrogen
[see Figure 2 of \citet{E90}].

{\it (iii) Outflow of enriched gas.} If the outflowing material
is uniform and well mixed, the effective yield will be
lower than the true yield and the abundance ratios, such
as N/O, will be unchanged.  Outflow of enriched gas can
be the result of at least two distinct physical processes: (1) 
``blow-out'' as a result of internal processes which inject
kinetic energy in excess of the binding energy of the galaxy
and (2) removal of the interstellar medium via external
processes such as ram pressure stripping.  In the
former case, the observed abundances are lower than
those predicted by closed box models because some of the 
enriched material is lost to the intergalactic medium. 
In the case of blow-out, one might expect a correlation between
effective yield and either total mass (or mass surface
density) or an anti-correlation with star formation rate.  However, it should
also be noted that despite their low mass, it appears
to be quite difficult to remove fully the interstellar
medium of typical dwarf irregular galaxies through internal processes
such as blow-out \citep[e.g.,][]{MF99,FT00,ST98}.

In the case of external triggers, the closed box model 
overpredicts the oxygen abundance because the gas mass fraction
is now underestimated due to loss of the interstellar
medium.  Such enriched gas outflows could account for the enrichment
patterns of dE/dSphs, which are both gas--poor and metal--poor
\citep[e.g.,][]{M98}.  Furthermore, studies of abundance
patterns of gas-rich galaxies in both cluster and field 
environments indicate that there are significant differences
in the effective yields, as would be expected
if the gas mass fractions of cluster galaxies are reduced
due to ram pressure stripping \citep[e.g.,][]{SKSZ96,LMR03}.  
At the same time, however, global scaling relations, such as 
the metallicity-luminosity relation, may not be affected significantly by
the environment \citep[e.g.,][]{LMR03}.

{\it (iv) Outflow of oxygen-rich gas from supernova winds.}  
In contrast to the previous case where the outflowing material is
assumed to be well mixed, it is also possible that the
outflowing gas is oxygen-rich since oxygen is formed in (short-lived)
high mass stars and nitrogen is formed in (longer-lived) intermediate
mass stars.  If the enriched gas is lost via supernova winds, the 
effective yields of nitrogen and oxygen may no longer be coupled
since recently enriched gas (oxygen-rich) is more likely to be ejected from
the system.  Differential outflow will result in higher N/O ratios
than expected since oxygen is preferentially depleted.
Nonetheless, even with differential outflow, nitrogen may also be 
depleted since the outflow is likely to include the surrounding
interstellar medium as well as the oxygen-rich ionized gas.
However, with typical star formation rates of $\sim$0.001 -- 0.01 
M$_{\odot}$ yr$^{-1}$ (surface rates of 0.0001 to 0.001 M$_{\odot}$ yr$^{-1}$ kpc$^{-2}$),
differential outflow may be less significant an issue for
quiescent dwarf irregular galaxies.  In contrast,
starbursting dwarf galaxies (e.g., blue compact dwarf galaxies)
will have periodic injections of substantial kinetic energy
when their high mass stars evolve, and thus may experience
substantial outflow of enriched material \citep[e.g., NGC 1569,][]{MKH02}.

{\it (v) Extended gas distributions: inefficient mixing.} 
The standard closed box model assumes that all of the gas is 
involved in the chemical evolution.  However, the HI distribution in
most dwarf irregular galaxies extends at least a factor
of two beyond the stellar disk \citep[e.g.,][]{BR97,SvvS02,vSH04}.
If the outlying gas is not enriched by on-going star formation
activity, the standard closed box model will underpredict the
elemental abundances because the enriched material is
mixing with a smaller fraction of the total gas mass \citep[e.g.,][]{GH84}.
{\bf Inefficient mixing is one of the few processes that result 
in effective yields which are higher than closed box yields.}
If the extended gas distribution is significantly lower abundance
than the inner gas, one might expect to see a correlation
between effective yield and size of the HI disk.  

Note that loss of pristine gas from the outer regions
due to ram pressure stripping has similar results
as described in (iii) since the overall gas mass
fraction will be reduced, thereby increasing the
predicted oxygen abundance.

{\it (vi) Time delay in delivery or mixing of products of stellar nucleosynthesis.}
Because oxygen and nitrogen are formed in different mass stars, there
may be a time delay between the observed enrichment of these two
elements \citep[e.g.,][]{EP78}.  
Oxygen formed in Type II supernovae will be released after $\sim$10 Myr
while nitrogen will be produced and released over a substantially
longer time period, $\sim$250 Myr, as the
intermediate mass stars evolve.  Thus, if the products of
stellar nucleosynthesis are mixed rapidly into the surrounding
medium, one expects to see a rapid decrease in the N/O ratio
when oxygen is released followed by a gradual increase in the
 N/O ratio as the intermediate mass stars have time to evolve
and release nitrogen into the surrounding medium. 
Thus, the N/O ratio may act as a `clock' marking the time since 
the last major star formation
episode \citep[e.g.,][]{EP78,SBK97,KS98,HEK00,SCM03}.  

\subsection{Scaling Relations}

\subsubsection{Metallicity-Luminosity Relation}
\begin{figure}
\includegraphics[angle=-90,width=17cm]{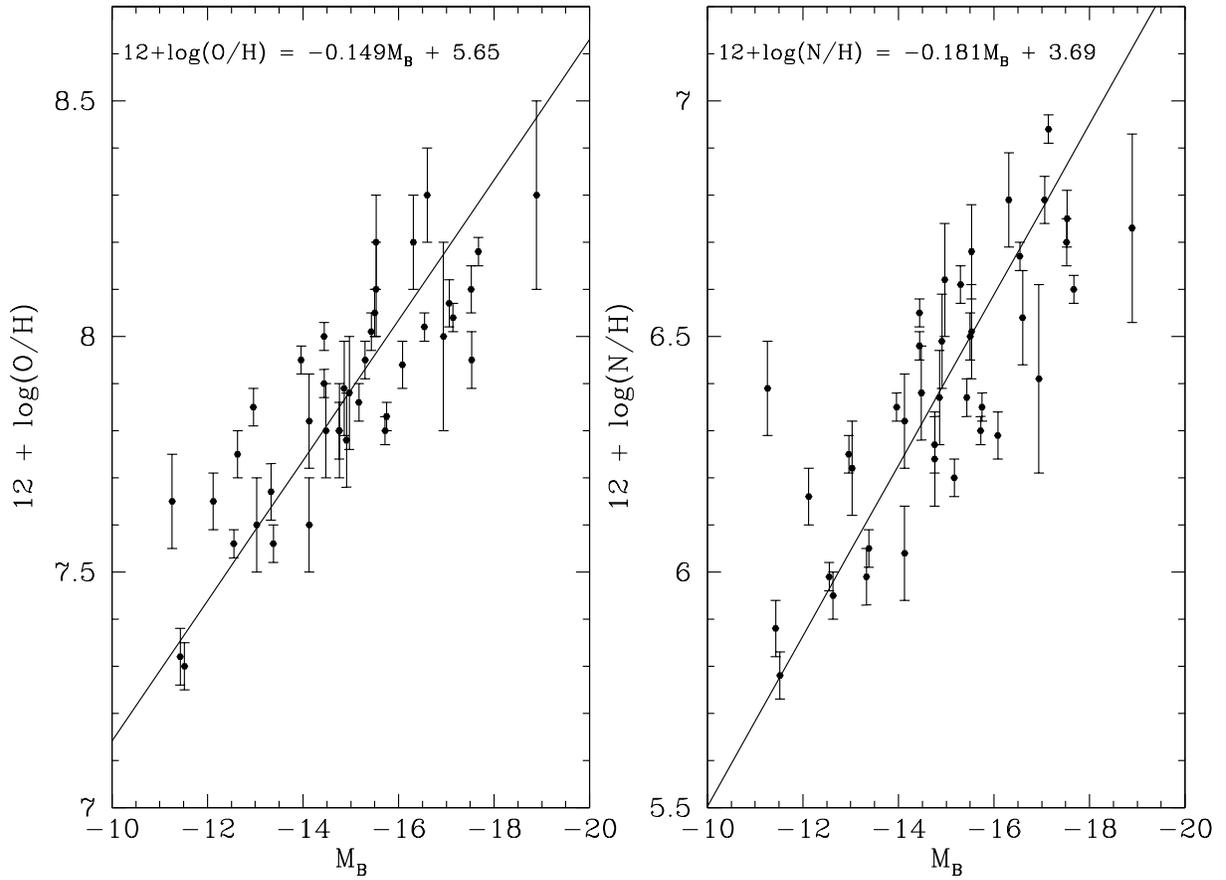}
\caption[]{
Metallicity-luminosity relationships for isolated dwarf irregular
galaxies (present sample; van Zee et al.\ 1997; van Zee et al.\ 2005).
Correlations between luminosity and both oxygen and nitrogen 
abundances are shown.  
\label{fig:mboh}
}
\end{figure}

Relationships between global elemental abundances of both oxygen and nitrogen
and absolute blue magnitude are shown in Figure \ref{fig:mboh}.  
The derived slope (-0.149 $\pm$ 0.011) and
intercept (5.65 $\pm$ 0.17) for the oxygen metallicity-luminosity relationship
for this sample are nearly identical to the relationship derived
from a recent literature compilation for galaxies within
5 Mpc \citep{vSH05} and to the field dwarf irregular
sample of \citet{LMKRS03}.  The relationship derived
from the 53000 galaxies in the SDSS data set is also consistent
with the present observations, although their overall trend
is steeper than that produced here \citep{T04}.

In addition to intrinsic scatter due to galactic evolution (either
chemical enrichment or luminosity evolution),
the scatter in the observed luminosity-metallicity
relationship may be due in part to the different oxygen abundance
calibrations (direct vs strong line) or to errors in 
the adopted distances.  First, as discussed in 
Section \ref{sec:strongline}, the strong line abundance calibration 
used here for some of the \hii regions may be systematically too 
high relative to direct abundance calculations.  However, since 
the global abundances are weighted more heavily by \hii regions 
with direct measurements, such effects should be relatively
small for this sample.  Systematic distance errors, which 
translate into incorrect absolute magnitudes, may be a more 
significant problem.  For example, all of the galaxies which 
lie on the upper envelope of the oxygen abundance-luminosity 
relation are nearby galaxies with distances determined from 
TRGB observations (HKK L14, GR 8, UGC 9128, UGC 8651, 
UGC 9240, and UGC 685, in order of increasing luminosity)
whereas the majority of distances in this study are estimated 
from Hubble flow models.  A slight shift in the Hubble flow 
distances (DM$_{err}$ $\sim$ 0.4 magnitudes, or a 20\% change in
the distance estimate) could bring these points 
into better agreement. 

The nitrogen abundance also has a strong correlation with
the luminosity.  The slope of the nitrogen abundance-luminosity 
correlation (-0.181 $\pm$ 0.014) is
consistent with the slope of the oxygen abundance
correlation.  However, this relationship has significantly
more scatter than the oxygen abundance-luminosity correlation (rms = 0.20 dex
for N/H vs 0.15 dex for O/H).  
As can be seen in Figure \ref{fig:no_plot}, several of the galaxies 
in this sample have relatively high nitrogen--to--oxygen abundance
ratios.  The increased scatter in the observed nitrogen--to--oxygen
abundance ratio at 12 + log(O/H) $>$ 7.9 introduces additional 
scatter into the nitrogen abundance-luminosity correlation since 
galaxies with high N/O ratios do not fall in any preferred location 
in the oxygen abundance-luminosity diagram.  We discuss
N/O ratios in more detail below (Section \ref{sec:no}).

The oxygen abundance-luminosity relation is the most commonly
adopted abundance scaling relation since both parameters
are easily obtained for large samples of galaxies \citep[e.g.,
the large spectroscopic samples from SDSS,][]{T04}.  However,
the underlying physical processes that drive these observed
scaling relations may be more directly tied to a mass-metallicity 
correlation since more massive galaxies are both more efficient in 
converting gas into stars (lower gas mass fractions at the present
epoch) and are able to retain a larger fraction of their processed 
metals due to their deeper gravitational potential wells.  Nonetheless, 
the oxygen-luminosity relation is used as a first-order
approximation since both stellar and dynamical masses are difficult 
to obtain: stellar mass estimates require knowledge of the underlying 
stellar population to determine a reliable mass--to--light ratio while dynamical
masses require additional (time intensive) kinematic observations.

In Figure \ref{fig:metlum_delta} we examine the residuals in the
metallicity-luminosity relation as a function of dynamical mass,
color, gas mass fraction, and surface brightness.  
While the dynamical mass estimates used here are approximate values 
based on global line widths, there is a suggestive trend that
more massive galaxies have lower oxygen abundances than
predicted by the metallicity-luminosity relation.
This trend is in the opposite sense as one would expect for
simple models of outflow or infall: the high mass
galaxies are on the low side of the residuals suggesting that
they are either underabundant or overluminous 
for their mass.  Thus, the scatter in the metallicity-luminosity 
relationship is not easily identified
as a result of either outflow (expected correlation with dynamical
mass) or evolution (expected anti-correlation with gas mass fraction).  
It is also important to note that there is no correlation
with central surface brightness, as one might expect if low
surface brightness galaxies are less evolved than high
surface brightness galaxies.

Finally, the fact that the nitrogen abundance
relation has more scatter than the oxygen abundance-luminosity
relation suggests that differential outflow (due to supernova winds,
for example) may not be a significant cause of the lower
effective yields and the high N/O ratios in dwarf irregular galaxies 
(Figure \ref{fig:yield}).  In some regards, one might expect the nitrogen 
abundance to be a better tracer of the chemical enrichment history of 
a galaxy because it is formed in intermediate mass stars and thus
is less likely to be lost from the system through outflow of enriched
gas.  In particular, differential outflow should introduce additional
scatter in the oxygen abundance-luminosity relation since its efficiency
will depend on a number of variables, such as the instantaneous star formation rate, 
mass surface density, and structure of the interstellar medium.  The tightness of the
oxygen luminosity-metallicity relation compared to the nitrogen relation 
suggests, however, that either the efficiency of differential outflow is 
common among all dwarf irregular galaxies or that its effects 
are negligible compared to evolutionary effects such as aging 
of the stellar population or time delay of delivery of enriched materials.

\subsubsection{N/O Ratios and Star Formation Histories}
\label{sec:no}

\begin{figure}
\epsscale{1.0}
\plotone{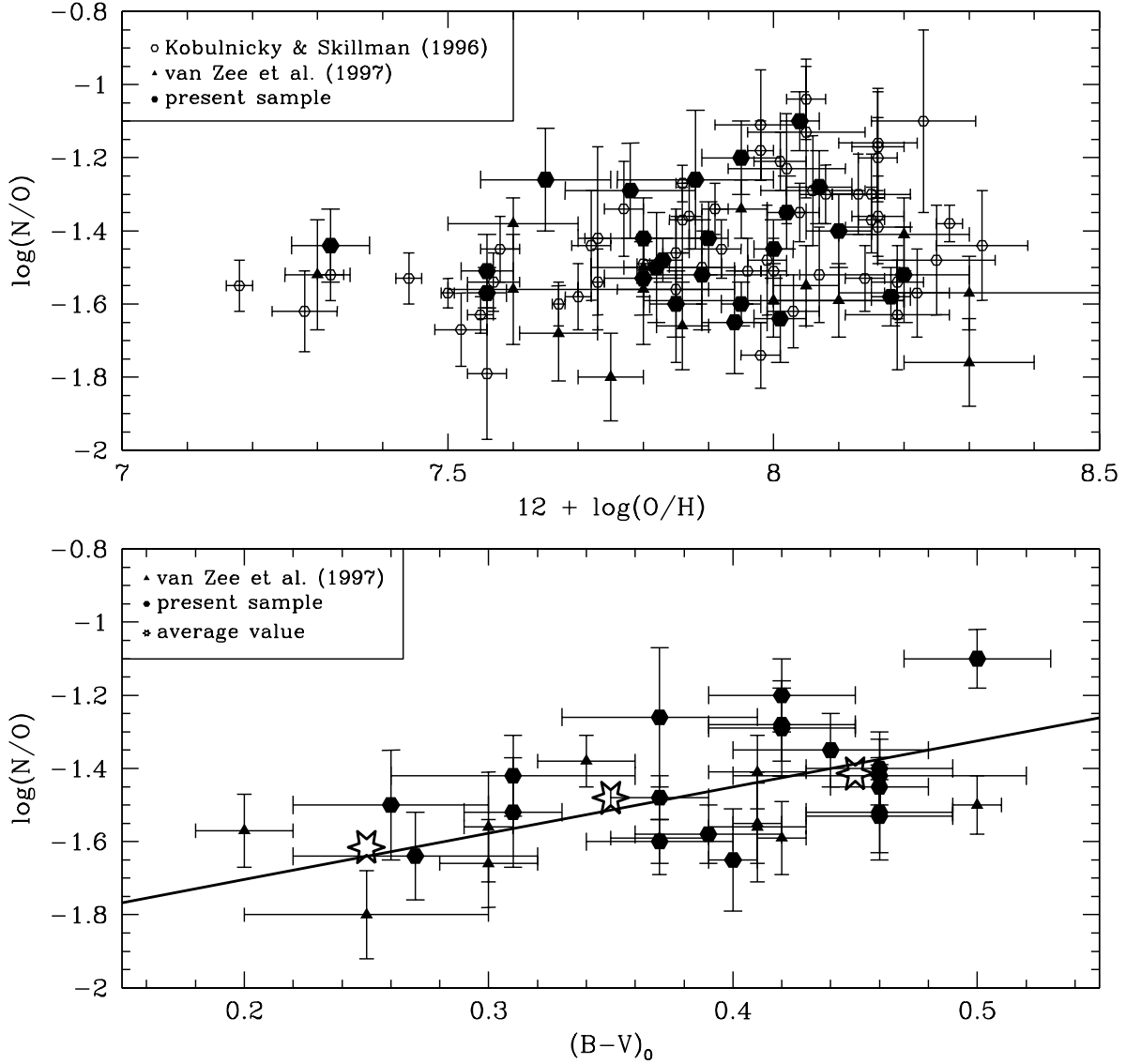}
\caption[]{
(Top) Global nitrogen and oxygen abundances for dwarf irregular galaxies.
The filled symbols are from the present (hexagons) and previous
(triangles; van Zee et al.\ 1997; 2005) studies of dwarf irregular galaxies.
The open symbols are from the literature compilation by 
Kobulnicky \& Skillman (1996); only galaxies without WR emission features
have been plotted.
(bottom) Correlation between nitrogen-to-oxygen ratio and
color of the underlying stellar population.  Redder
galaxies have higher N/O ratios as would be expected
for a time delay between the release of oxygen and
nitrogen.
\label{fig:no_plot}
}
\end{figure}

\begin{figure}
\plotone{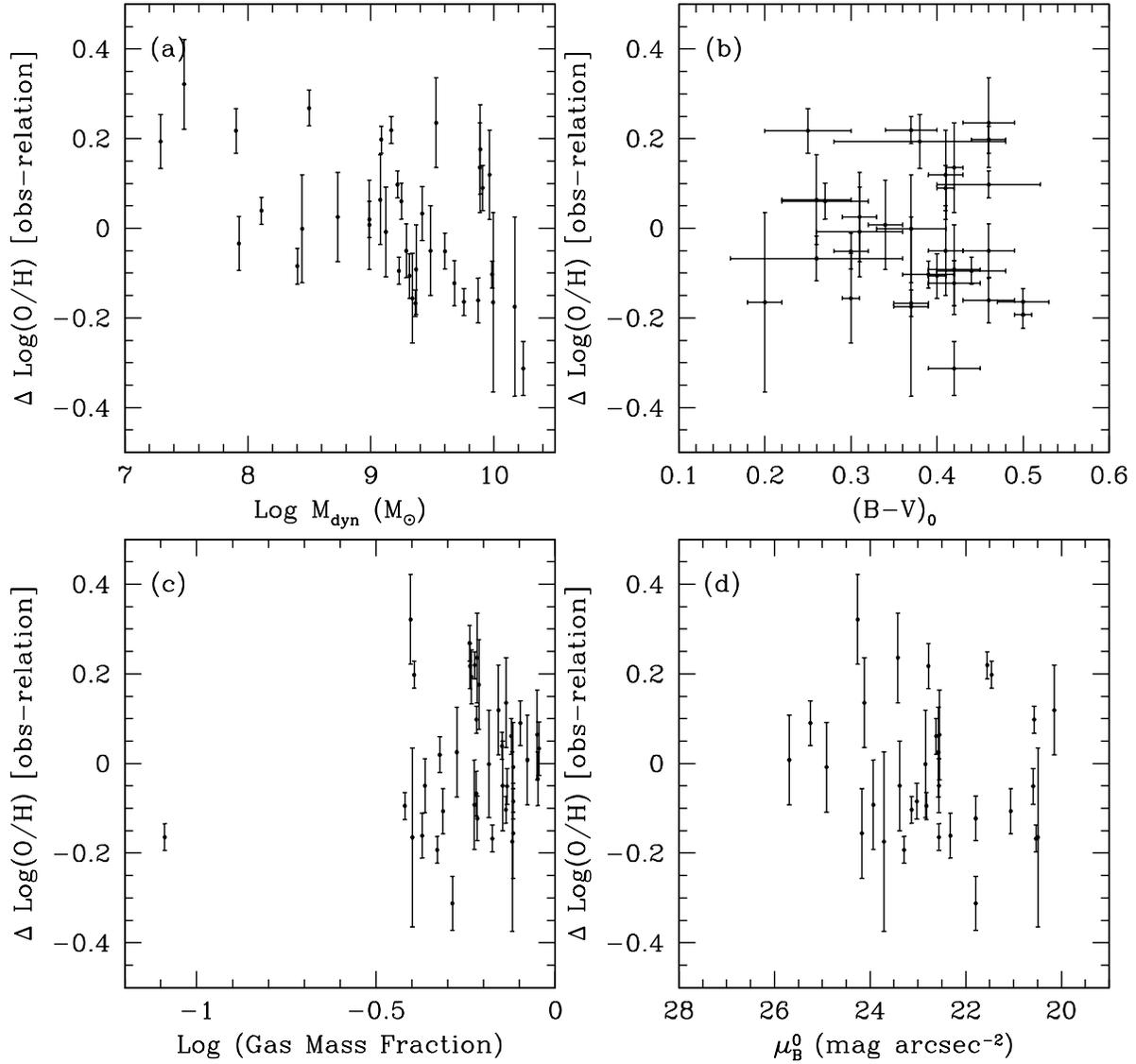}
\caption[]{
Residuals in the metallicity-luminosity relation plotted
as a function of other global galaxy parameters: (a)
dynamical mass, (b) color of the underlying stellar population,
(c) gas mass fraction, and (d) central surface brightness.
\label{fig:metlum_delta}
}
\end{figure}

The scatter in the N/O ratio at intermediate metallicities
(Figure \ref{fig:no_plot})
has been attributed to several possible causes: (1)
a time delay between release of oxygen and nitrogen;
(2) additional production of secondary nitrogen; and
(3) differential outflow of oxygen-rich gas.
First, as discussed above (Section 4.3.1), differential 
outflow is unlikely to be the primary cause of the high N/O ratios
in this sample because such a process predicts a tighter metallicity-luminosity
relation for nitrogen (minimal outflow) than oxygen (outflow with
variable efficiencies).  Thus, while most of the galaxies
in this sample appear to be open boxes (Section 4.1), the relative abundance
ratios appear to be indicative of evolutionary effects rather than
galactic processes (Section 4.2).  Rather, the scatter in the observed
N/O ratios may be due either to varied star formation rate histories
or to the increasing production of secondary nitrogen at intermediate metallicities.
For clarity,
we initially discuss time delay issues for 
primary nitrogen alone (1) and then broaden the discussion
to include the effects of additional synthetic pathways (2) on the
observed abundance ratios.

We begin with a brief summary of the sources of primary 
and secondary nitrogen.
At high metallicities, the nitrogen yield is linearly
related to the initial metallicity (C and O) of the
star; in other words, nitrogen is a secondary
element and metal--rich stars will produce more nitrogen
than metal--poor stars.  Secondary nitrogen is produced
in the core of intermediate mass stars during the CNO cycle \citep[e.g.,][]{C83}.
At low metallicities, however, nitrogen appears to be
a primary element, i.e., produced only out of the original
hydrogen in the star. Primary nitrogen may be produced
in intermediate mass stars in which the C and O produced in the 
helium burning core is mixed into a hydrogen burning shell \citep[e.g.,][]{RV81}.
Primary nitrogen may also be produced in low metallicity
massive stars via convective overshoot \citep[e.g.,][]{TWW95,WW95}.
If primary nitrogen is produced in the same stars as oxygen,
then the enrichment timescales for N and O are identical
and the observed N/O ratio (for primary N) should be
constant, regardless of overall oxygen abundance or
age of the stellar population.  However, if intermediate
mass stars are the primary source of nitrogen, then
there may be a significant time delay between oxygen
enrichment and nitrogen enrichment; the length of the
delay will depend both on stellar evolution timescales and on the
cooling and mixing time of the enriched gas.

If the mixing timescales are short relative to stellar evolution, the scatter in the 
N/O vs O/H diagram may be a natural result of differing 
star formation histories. For example, a galaxy with
a constant star formation rate will have a lower net
nitrogen-to-oxygen yield than a galaxy with a declining
star formation rate because more oxygen will have been
released to the ISM due to the on--going star formation
activity.  Figure \ref{fig:no_plot} shows the correlation
in observed N/O ratio and the color of the underlying
stellar population.  While the color range is not large,
a least-squares fit to the data yields:
\begin{equation}
{\rm log(N/O)} = 1.26 {\rm (B-V)_0} - 1.96.
\end{equation}
Also shown in Figure \ref{fig:no_plot} are the
mean values for 3 color bins;
the mean log(N/O) is --1.61 $\pm$ 0.14 for
galaxies with 0.2 $<$ \bv $<$ 0.3, --1.48 $\pm$ 0.11 for
galaxies with 0.3 $<$ \bv $<$ 0.4, and --1.41 $\pm$ 0.13 for
galaxies with 0.4 $<$ \bv $<$ 0.5.
 While degeneracy between
age and metallicity of the dominant stellar population
may introduce subtle effects in this diagram, the correlation
between color and N/O ratio indicates that the high N/O ratios 
are a result of declining star formation rates.
Note, however, that color is independent
of dynamical mass and luminosity for this sample of galaxies. 
The correlation is well defined in Figure \ref{fig:no_plot}
in part because the present sample includes quiescent
dwarf irregular galaxies with star formation histories
that are well modelled by simple changes in the star formation
rate.  It is important to note that this scenario does not
require star formation to occur in periodic starbursts.
Rather, as long as the recent star formation activity is less 
than the past star formation rate, the delayed release of nitrogen 
from the aggregate intermediate mass stellar population will
slowly increase the N/O ratio since it is not balanced
by additional production of oxygen in high mass stars.
While none exist in the present sample,
it is also possible to create blue galaxies with high N/O
ratios if one considers more complicated star formation
histories; for example, if a major starburst episode begins after a 
long period of quiescence, the elemental abundances will
be indicative of the previous enrichment history while the
colors are indicative of the present star formation activity.
Such pathological cases exist \citep[e.g., NGC 5253,][]{KSRWR97}.
Nonetheless, it is clear that high N/O ratios are correlated
with redder systems and decreasing star formation rates
for this representative sample of dwarf irregular galaxies.

If high N/O ratios are a natural result of the combined influences of
varied star formation histories and delayed delivery of nitrogen 
from intermediate mass stars,
then we would expect to see a range of N/O ratios
at all metallicities since any red, low luminosity galaxy
should have high N/O.  However, to date, there are
no known examples of extremely low metallicity galaxies with high N/O
ratios.  Rather, the N/O ratio appears to have minimal
scatter until an oxygen abundance of 12 + log O/H $\sim$ 7.9.
In fact, \citet{IT99} argue that the extremely small scatter 
in the N/O ratios of starbursting dwarf galaxies indicates that
primary nitrogen is produced in massive stars (i.e., in the same
stars that produce oxygen) and that extremely low metallicity galaxies are young
(i.e., there has not been sufficient time since the onset of star formation
for intermediate mass stars to evolve and release additional nitrogen).
Since the low metallicity regime is precisely where we expect to
be most sensitive to small increases in the nitrogen yield from intermediate mass stars,
these observational results are difficult to explain in the context
of normal star formation histories and chemical enrichment scenarios.
Nonetheless, there are severe selection effects that bias current
samples against red, low metallicity galaxies since the star formation
rates of these galaxies will be extremely low; most metallicity studies
(including this one) exclude galaxies with extremely faint HII regions
since interpretation of the line ratios may be ambiguous (see Section \ref{sec:strongline}).
Nonetheless, observations of oxygen and nitrogen abundances
in such low luminosity galaxies may be necessary to understand
the true scatter in the N/O ratio, and thus to understand
fully the chemical enrichment history of dwarf irregular galaxies.

Alternatively, the increased scatter in the N/O ratio may be 
a result of the increased importance of secondary nitrogen
production as the metallicity of the system increases.  Since 
secondary nitrogen is formed in intermediate mass stars, this 
scenario incorporates all of the aspects described above, but 
also introduces an additional free parameter: the effective 
yield of secondary nitrogen.  Aside from the first generation of 
stars, any subsequent stellar population will include an increasing 
fraction of processed materials, and thus both primary and secondary 
nitrogen will be produced in intermediate mass stars.  The point of 
inflection in the N/O plot at an oxygen abundance of 12 + log(O/H) 
$\sim$ 8.3 is usually interpreted as the transition regime
between secondary and primary nitrogen \citep[e.g.,][]{VE93,vSHOB98,HEK00}.
Since secondary nitrogen is formed in intermediate mass
stars, a decreasing star formation rate will yield
a high N/O ratio as more secondary nitrogen is slowly released 
into the interstellar medium without a corresponding increase 
in the oxygen abundance.  Since the relative yields of primary 
and secondary nitrogen are metallicity dependent, this scenario 
predicts that the dispersion in N/O ratio should increase with 
oxygen abundance in the metallicity range where both primary and 
secondary processes are significant contributors to the nitrogen
abundance.  

Finally, as illustrated by \citet{KH05}, some of the scatter
in the N/O ratio at intermediate metallicities may also be a 
result of global gas flows.  As discussed in Section 4.2,
infall or outflow of uniform, well mixed, material will not
affect the observed N/O ratios.  However, \citet{KH05} point
out that infall of pristine gas onto a system that has
already reached the secondary nitrogen regime may
drive the system into the region of the abundance diagram 
populated by dwarf irregular galaxies since the oxygen
abundance will be diluted but the N/O ratio remains unchanged.
Assuming a reasonable mass distribution of such infall events
also leads to a metallicity dependence in the dispersion
of the N/O ratio.  Thus, this scenario predicts that the observed 
scatter in the N/O ratio should be correlated with gas richness
as infalling pristine gas will increase the gas mass fraction.
However, the actual trend is in the opposite sense; galaxies
with high gas mass fractions tend to have lower N/O ratios,
while the high N/O ratios are found in systems with low gas
mass fractions.  In fact, all of the `closed box' galaxies in 
this sample are on the lower envelope of observed N/O ratios.  
While this result may simply reflect that gas-rich systems are more
likely to have continuous star formation activity,
it argues against producing the observed scatter in N/O
by diluting systems which have reached the secondary nitrogen
regime.

Chemical evolution models which trace gas infall, outflow,
star formation history, and enrichment patterns of 
all the elements may be necessary to disentangle fully the
relative importance of time delays, secondary nitrogen
production, and gas flows.  At this time, several
chemical enrichment models are able to reproduce the global trends in
the  nitrogen-to-oxygen abundance ratio at both low 
and high metallicities by invoking several different
physical processes \citep[e.g.,][]{CCP99,LSLP01,MC02,KH05}.
A reasonable goal for future chemical evolution models is to produce
both self-consistent star formation and chemical enrichment
histories and to reproduce the scatter in observed abundance ratios.

\subsection{Deviation from Closed Box Chemical Evolution}

\begin{figure}
\plotone{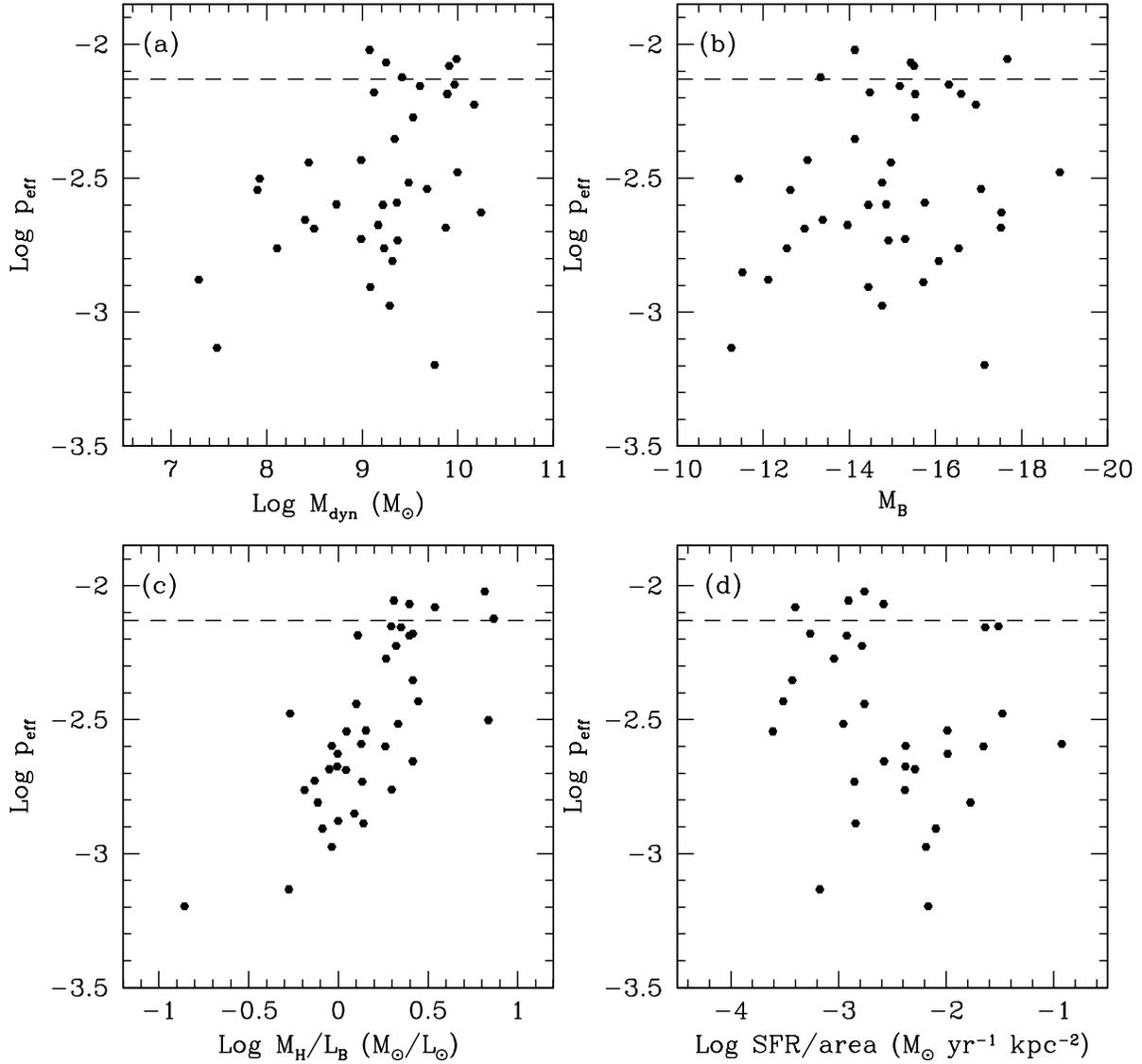}
\caption[]{
The effective yield plotted as a function of global
galaxy parameters: (a) dynamical mass, (b) absolute
blue magnitude, (c) gas richness (M$_{\rm H}$/L$_{\rm B}$),
and (d) surface star formation rate (star formation rate/area).
The closed box yield is denoted by the dashed line in
each plot.  The effective yield correlates strongly 
with gas richness, but no other global trends are evident.
\label{fig:delta_closed}
}
\end{figure}

As discussed in Section 4.1, a handful of galaxies in
this sample appear to be consistent with simple closed
box chemical evolution; the majority, however, have
lower effective yields than those predicted by closed box evolution
and thus must have experienced significant gas infall or outflow. 
 In this section, we examine global properties of dwarf irregular
galaxies to investigate whether there are significant
differences between `open' and `closed' systems.
The effective yield is shown as a function of dynamical
mass, absolute magnitude, M$_{\rm H}$/L$_{\rm B}$, and
surface star formation rate (SFR/area) in Figure \ref{fig:delta_closed}.

As discussed in Section 4.2, if the oxygen abundance is
diluted significantly by infall of pristine gas, we expect
to see a net anti-correlation between the effective yield and 
gas-richness since the infalling gas increases the total
gas mass fraction of the system.  However, as indicated in 
Figure \ref{fig:delta_closed}c, this prediction is opposite to
the trend observed.  Rather, gas-rich galaxies appear to be 
closed systems while those with lower M$_{\rm H}$/L$_{\rm B}$ have 
lower effective yields.  Thus, as might be expected given
the dearth of free floating HI clouds in the local universe,
infall of pristine gas does not appear to be the dominant physical 
process that lowers the effective yields of dwarf irregular
galaxies. 

Discussions about the significance of gas infall in the
chemical evolution of dwarf irregular galaxies are often
complicated by the fact that many dIs are surrounded by 
large HI disks \citep[e.g.,][]{BR97,SvvS02,vSH04}.
If the extended gas envelope has not been involved in
the star formation process, this may be considered a potential
source for infalling pristine gas \citep[e.g., DDO 154 = UGC 8024,][]{KS01}.
However, in the context of this discussion, these
extended gas disks should not be considered a source
of pristine material since the closed box model presented here
includes the {\it global} gas content.  In other words,
the extended gas disk has already been counted in the gas
mass fraction; if only the gas associated with the stellar
disk is considered, the effective yields are even lower
than those presented here.  Further analysis of the
chemical enrichment process which includes detailed
knowledge of the gas distribution and
localized gas content (coincident with the stellar distribution,
for example) is planned for a future paper.

The possibility that gas loss (outflow) drives the observed 
correlation between luminosity and metallicity 
has been considered in detail by both \citet{G02} and
\citet{T04}.   Both of these studies indicate that
massive gas-rich galaxies (spirals) are consistent with closed
box chemical evolution while low mass galaxies appear
to experience significant outflow.  However, the galaxies
in the present sample do not appear to confirm these
trends (Figure \ref{fig:delta_closed}a).  Rather, 
the only apparent correlation between effective yield
and global galaxy properties is with gas-richness; none
of the other global parameters, including dynamical mass, 
absolute magnitude, current star formation activity, 
and surface brightness, have significant correlations with 
effective yield.  Thus, while enriched gas outflow is a natural 
explanation for low effective yields in low mass galaxies, 
and some dwarf galaxies appear to have lost a significant
fraction of their metals, there
do not appear to be simple physical processes that dictate
whether a galaxy will be an open or closed box.

\section{Conclusions}
We present the results of optical spectroscopy of 67 \hii regions in
21 dwarf irregular galaxies.  Elemental abundances of oxygen,
nitrogen, neon, sulfur, and argon are derived and considered
in the context of simple chemical evolution models.
The major results are summarized below.

(1) Oxygen abundances are obtained via direct detection of
the temperature sensitive lines for 25 \hii regions (14 galaxies).
The direct abundance calculations are compared with strong line 
abundance calibrations. The
photoionization calibration of \citet{M91} yields a statistical
offset of 0.07 dex, but the residuals have no apparent systematic trends.
In contrast, while the empirical calibration of \citet{P00} yields a
 negligible statistical offset for these \hii regions, the residuals 
are correlated with ionization parameter.  Thus, these observations 
indicate that oxygen abundances will be overestimated by the p-method 
for \hii regions with low ionization parameters.

(2) H II region abundances are calculated at a range of radii for 12 of 
the 21 galaxies.  In general, oxygen abundances derived from 
observations of multiple \hii regions within the same galaxy are 
quite similar, indicating that the interstellar medium is well mixed.  

(3) Global elemental abundances for this sample of 
dwarf irregular galaxies are similar to those observed in
other low mass dwarf galaxies.  The mean log(N/O) abundance
ratio is -1.43 $\pm$ 0.15, the mean log(Ne/O) abundance ratio is -0.79 $\pm$ 0.09,
the mean log(S/O) abundance ratio is -1.53 $\pm$ 0.09, and
the mean log(Ar/O) abundance ratio is -2.19 $\pm$ 0.09.

(4) The luminosity-metallicity relation for this sample
of isolated dwarf irregular galaxies is: 12 + log(O/H) = 
-0.149 M$_{\rm B}$ + 5.65.  The nitrogen abundance-luminosity
relation is: 12 + log(N/H) = -0.181 M$_{\rm B}$ + 3.69.
The nitrogen relation has significantly more
scatter than the oxygen metallicity-luminosity relation.

(5) A correlation is found between the observed nitrogen--to--oxygen 
ratio and the color of the underlying stellar population in
the sense that redder dwarf irregulars have higher N/O ratios than blue
dwarf irregulars.  This correlation is interpreted in
the context of delayed release of nitrogen and varied
star formation histories.  The N/O ratio will gradually
increase in galaxies with declining star
formation rates as the newly released nitrogen from
evolved intermediate mass stars will  no longer be
balanced by the influx of additional oxygen from
recently formed high mass stars.

(6) Global oxygen and nitrogen abundances are examined in 
the context of open and closed box chemical evolution models.
While several galaxies are consistent with closed box chemical
evolution, the majority of this sample have an effective
yield $\sim$1/4 of the expected yield for a Salpeter IMF
and constant star formation rate.  However, the effective
yield is not correlated with global galaxy parameters
such as dynamical mass, absolute magnitude, star formation
rate, or surface brightness.  Rather, the effective
yield is strongly correlated with M$_{\rm H}$/L$_{\rm B}$
in the sense that gas--rich galaxies are more likely
to be closed boxes.

\acknowledgements
We thank Evan Skillman for numerous discussions about dwarf
galaxies and metallicity evolution.
We thank Dick Henry and Henry Lee for helpful comments on
early versions of this paper.
This research has made use of the NASA/IPAC Extragalactic Database (NED)
which is operated by the Jet Propulsion Laboratory, California Institute
of Technology, under contract with the National Aeronautics and Space
Administration.
LvZ acknowledges partial support from the 
Herzberg Institute of Astrophysics and the National Research Council of Canada;
LvZ also acknowledges partial support from Indiana University.
MPH has been supported by NSF grants AST-9900695 and AST-0307396.

\clearpage

\begin{deluxetable}{lrrccccccccccc}
\tabletypesize{\tiny}
\tabcolsep 2pt
\tablewidth{0pt}
\tablecaption{Global Galaxy Parameters\label{tab:gals}}
\tablehead{
\colhead{} & \colhead{RA} &\colhead{Dec}& \colhead{Morph.} & \colhead{Distance}& \colhead{R$_{25}$} & \colhead{}& \colhead{} & \colhead{} & \colhead{}& \colhead{} & \colhead{12 + } & \colhead{}\\[.2ex]
 \colhead{Galaxy}& \colhead{(2000)} & \colhead{(2000)} & \colhead{Type}& \colhead{[Mpc]}&\colhead{arcsec (kpc)}& \colhead{M$_{\rm B}$} & \colhead{(B-V)$_0$} & \colhead{M$_{\rm H}$/L$_{\rm B}$} & \colhead{$\mu_B^0$}& \colhead{log(M$_{\rm dyn}$)} &\colhead{log(O/H)} & \colhead{log(N/O)}
}
\startdata
UGC 12894    & 00 00 22.2 & 39 29 46.6 & Im    &  7.9 & 21 (0.8) & -13.38 &       \nodata   & 2.5 & 23.0 & 8.40 & 7.56 $\pm$ 0.04 & -1.51 $\pm$ 0.10 \\
UGC   290    & 00 29 07.9 & 15 54 02.9 & Sdm   & 12.7 & 42 (2.6) & -14.48 & 0.31 $\pm$ 0.05 & 2.6 & 24.9 & 9.12 & 7.80 $\pm$ 0.10 & -1.42 $\pm$ 0.11 \\
UGC   685    & 01 07 22.4 & 16 41 04.3 & SAm   & 4.79 & 44 (1.0) & -14.44 & 0.46 $\pm$ 0.02 & 0.8 & 21.5 & 9.08 & 8.00 $\pm$ 0.03 & -1.45 $\pm$ 0.08 \\
UGC  1104    & 01 32 42.4 & 18 19 00.9 & Im    & 11.1 & 32 (1.7) & -16.08 & 0.40 $\pm$ 0.01 & 0.8 & 21.1 & 9.31 & 7.94 $\pm$ 0.05 & -1.65 $\pm$ 0.14 \\
UGC  1175    & 01 39 56.5 & 11 05 46.9 & Sm:   & 11.3 & 18 (1.0) & -14.13 & 0.26 $\pm$ 0.04 & 6.6 & 22.5 & 9.08 & 7.82 $\pm$ 0.10 & -1.50 $\pm$ 0.15 \\
UGC  1281    & 01 49 31.4 & 32 35 16.6 & SdM   &  4.6 &136 (3.0) & -14.91 & 0.42 $\pm$ 0.03 & 1.4 & 23.9 & 9.37 & 7.78 $\pm$ 0.10 & -1.29 $\pm$ 0.13 \\
HKK97 L14    & 02 00 10.2 & 28 49 46.5 & Irr   &  4.7 & 12 (0.3) & -11.26 &       \nodata   & 0.5 & 24.3 & 7.48 & 7.65 $\pm$ 0.10 & -1.26 $\pm$ 0.14 \\
UGC  2023    & 02 33 18.6 & 33 29 28.7 & Im:   & 10.2 & 44 (2.2) & -16.54 & 0.44 $\pm$ 0.04 & 0.6 & 22.8 & 9.23 & 8.02 $\pm$ 0.03 & -1.35 $\pm$ 0.10 \\
UGC  3647    & 07 04 50.0 & 56 31 09.8 & IBm   & 19.7 & 40 (3.8) & -17.06 & 0.42 $\pm$ 0.03 & 1.4 & 21.8 & 9.68 & 8.07 $\pm$ 0.05 & -1.28 $\pm$ 0.10 \\
UGC  3672    & 07 06 27.3 & 30 19 21.1 & Im    & 12.7 & 48 (3.0) & -15.43 & 0.27 $\pm$ 0.05 & 2.5 & 22.6 & 9.25 & 8.01 $\pm$ 0.04 & -1.64 $\pm$ 0.12 \\
UGC  4117    & 07 57 25.9 & 35 56 24.8 & IBm   & 10.0 & 37 (1.8) & -14.86 & 0.31 $\pm$ 0.02 & 0.9 & 22.6 & 8.73 & 7.89 $\pm$ 0.10 & -1.52 $\pm$ 0.15 \\
UGC  4483    & 08 37 03.0 & 69 46 36.3 & Im    & 3.21 & 34 (0.5) & -12.55 &       \nodata   & 2.0 &\nodata& 8.11& 7.56 $\pm$ 0.03 & -1.57 $\pm$ 0.07 \\
CGCG 007-025 & 09 44 02.1 &-00 38 32.6 & Sm    & 17.8 & 15 (1.3) & -15.75 & 0.37 $\pm$ 0.02 & 1.3 & 20.5 & 9.36 & 7.83 $\pm$ 0.03 & -1.48 $\pm$ 0.06 \\
UGC  5288    & 09 51 17.0 & 07 49 40.0 & Sdm:  &  5.3 & 37 (1.0) & -14.44 & 0.46 $\pm$ 0.06 & 1.8 & 20.6 & 9.22 & 7.90 $\pm$ 0.03 & -1.42 $\pm$ 0.06 \\
UGCA 292     & 12 38 39.7 & 32 45 49.3 & ImIV-V&  3.1 & 30 (0.4) & -11.43 & 0.08 $\pm$ 0.10 & 6.9 &\nodata& 7.93& 7.32 $\pm$ 0.06 & -1.44 $\pm$ 0.10 \\
UGC  8651    & 13 39 53.1 & 40 44 20.4 & Im    & 3.01 & 69 (1.0) & -12.96 &       \nodata   & 1.1 &\nodata& 8.50& 7.85 $\pm$ 0.04 & -1.60 $\pm$ 0.09 \\
UGC  9240    & 14 24 43.8 & 44 31 33.7 & IAm   & 2.79 & 57 (0.8) & -13.96 & 0.37 $\pm$ 0.03 & 1.0 & 21.6 & 9.17 & 7.95 $\pm$ 0.03 & -1.60 $\pm$ 0.06 \\
UGC  9992    & 15 41 47.7 & 67 15 13.9 & Im    &  8.6 & 32 (1.3) & -14.97 & 0.37 $\pm$ 0.04 & 1.2 & 22.8 & 8.44 & 7.88 $\pm$ 0.12 & -1.26 $\pm$ 0.19 \\
UGC 10445    & 16 33 47.6 & 28 59 05.1 & SBc   & 15.1 & 60 (4.4) & -17.53 & 0.42 $\pm$ 0.03 & 1.0 & 21.8 &10.24 & 7.95 $\pm$ 0.06 & -1.20 $\pm$ 0.10 \\
UGC 11755    & 21 29 59.6 & 02 24 50.8 & BCD/E & 18.0 & 27 (2.4) & -17.14 & 0.50 $\pm$ 0.03 & 0.1 & 22.6 & 9.76 & 8.04 $\pm$ 0.03 & -1.10 $\pm$ 0.08 \\
UGC 12713    & 23 38 14.4 & 30 42 30.6 & S0/a  &  7.5 & 31 (1.1) & -14.76 & 0.46 $\pm$ 0.03 & 0.9 & 22.6 & 9.29 & 7.80 $\pm$ 0.06 & -1.53 $\pm$ 0.10 \\
\enddata
\tablecomments{Distances are calculated from
systemic velocities using a Virgocentric infall model and
an H$_0$ of 75 km s$^{-1}$ Mpc$^{-1}$ except for
UGC 685 (Ma\'iz-Apell\'aniz et al.\ 2002),
HKK97 L14 (Karachentzev et al.\ 2004),
UGC 4483 (Dolphin et al.\ 2001),
and UGC 8651 and UGC 9240 (Karachentzev et al.\ 2002),
which have TRGB distances in the literature.
Colors, magnitudes, optical size, and M$_H$/L$_B$ are from van Zee (2000).
Global oxygen and nitrogen abundances are from the
present study except for UGC 3672, for which more accurate
abundances are available in the literature (van Zee et al.\ 1997).  }
\end{deluxetable}

\clearpage

\begin{deluxetable}{lrrrrr}
\tabletypesize{\footnotesize}
\tablewidth{0pt}
\tablecaption{Observing Log\label{tab:slits}}
\tablehead{
\colhead{Slit } & \colhead{RA} &\colhead{Dec}& \colhead{PA} & \colhead{}& \colhead{T$_{\rm int}$} \\
 \colhead{Position}& \colhead{(2000)} & \colhead{(2000)} & \colhead{[deg]}& \colhead{Run}& \colhead{[sec]}
}
\startdata
UGC 12894-A & 00 00 23.8 &   39 29 43 &  89 & Sep00 & 3 $\times$ 1200 \\
UGC   290-A & 00 29 08.3 &   15 53 57 & 135 & Sep00 & 3 $\times$ 1200 \\
UGC   685-A & 01 07 22.8 &   16 40 53 & 334 & Dec99 & 3 $\times$ 1200 \\
UGC   685-B & 01 07 23.2 &   16 41 00 &  41 & Dec99 & 3 $\times$ 1200 \\
UGC  1104-A & 01 32 42.6 &   18 19 08 & 170 & Dec99 & 3 $\times$ 1200 \\
UGC  1104-B & 01 32 42.7 &   18 19 00 & 109 & Dec99 & 3 $\times$ 1200 \\
UGC  1175-A & 01 39 56.5 &   11 05 40 & 163 & Sep00 & 3 $\times$ 1200 \\
UGC  1281-A & 01 49 32.1 &   32 35 30 &  90 & Dec99 & 2 $\times$ 1200 \\
HKK97 L14 -A& 02 00 10.4 &   28 49 43 &  70 & Dec99 & 3 $\times$ 1200 \\
UGC  2023-A & 02 33 15.7 &   33 29 19 &  80 & Dec99 & 1 $\times$ 1200 \\
UGC  2023-B & 02 33 17.9 &   33 29 54 &  84 & Dec99 & 3 $\times$ 1200 \\
UGC  2023-C & 02 33 18.1 &   33 29 51 &  93 & Dec99 & 3 $\times$ 1200 \\
UGC  3647-A & 07 04 50.6 &   56 31 14 &  26 & Dec99 & 1 $\times$ 1200 \\
UGC  3672-A & 07 06 23.3 &   30 20 52 & 106 & Sep00 & 2 $\times$ 1200 \\
UGC  4117-A & 07 57 25.0 &   35 56 42 &  90 & Mar00 & 5 $\times$ 1200 \\
UGC  4117-B & 07 57 25.8 &   35 56 33 & 110 & Dec99 & 5 $\times$ 1200 \\
UGC  4483-A & 08 37 02.8 &   69 46 52 & 175 & Dec99 & 1 $\times$ 1200 \\
CGCG 007-025-A& 09 44 01.8&--00 38 32 & 336 & Jan97 & 1 $\times$  900 \\
UGC  5288-A & 09 51 16.5 &   07 49 50 & 146 & Dec99 & 3 $\times$ 1200 \\
UGC  5288-B & 09 51 17.4 &   07 49 27 &   9 & Feb00 & 3 $\times$ 1200 \\
UGC  5288-C & 09 51 18.7 &   07 49 07 & 175 & Dec99 & 5 $\times$ 1200 \\
UGCA 292-C  & 12 38 40.0 &   32 46 02 & 112 & Apr02 & 9 $\times$ 1200 \\
UGCA 292-D  & 12 38 42.5 &   32 45 46 &  90 & Apr02 & 4 $\times$ 1200 \\
UGC  8651-A & 13 39 57.0 &   40 44 56 &  90 & Apr02 & 3 $\times$ 1200 \\
UGC  9240-A & 14 24 44.2 &   44 30 57 &  90 & Apr01 & 3 $\times$ 1200 \\
UGC  9992-A & 15 41 50.0 &   67 15 02 & 165 & Apr02 & 4 $\times$ 1200 \\
UGC 10445-A & 16 33 45.6 &   28 59 03 &  61 & Apr01 & 2 $\times$ 1200 \\
UGC 10445-B & 16 33 46.1 &   28 59 53 & 110 & Apr01 & 3 $\times$ 1200 \\
UGC 11755-A & 21 29 59.4 &   02 24 54 & 149 & Sep00 & 2 $\times$ 1200 \\
UGC 12713-A & 23 38 14.3 &   30 42 28 &  90 & Dec99 & 3 $\times$ 1200 \\
UGC 12713-B & 23 38 14.7 &   30 42 35 &  62 & Dec99 & 1 $\times$ 1200
\enddata
\end{deluxetable}

\clearpage

\begin{deluxetable}{lcccccccccccccc}
\tabletypesize{\tiny}
\tabcolsep 2pt
\rotate
\tablewidth{0pt}
\tablecaption{HII Region Line Strengths\label{tab:flux}}
\tablehead{
\colhead{Galaxy/} & \colhead{Location} &\colhead{[OII]}& \colhead{[NeIII]} & \colhead{[OIII]} & \colhead{[OI]} & \colhead{[SIII]} & \colhead{H$\alpha$} & \colhead{[NII]} & \colhead{He} & \colhead{[SII]} & \colhead{[ArIII]} & \colhead{C$_{H\beta}$} & \colhead{EW} & \colhead{H$\beta$} \\
 \colhead{Slit}& \colhead{EW  NS } & \colhead{3727+3729} & \colhead{3869}& \colhead{4959+5007} &\colhead{6300}& \colhead{6312}&\colhead{6563} & \colhead{6548+6584} & \colhead{6678} & \colhead{6717+6731} &\colhead{7136}
& \colhead{} & \colhead{H$\beta$} & \colhead{[10$^{-15}$]}
}
\startdata
U12894/A & -006 -005& 1.549$\pm$0.095& 0.286$\pm$0.039& 3.519$\pm$0.121& \nodata& \nodata& 2.752$\pm$0.157& 0.080$\pm$0.013& 0.043$\pm$0.009& 0.182$\pm$0.015& \nodata& 0.35$\pm$ 0.06&  63&  0.29\\
U12894/A & +004 -004& 1.845$\pm$0.105&   \nodata      & 3.044$\pm$0.102& \nodata& \nodata& 2.831$\pm$0.156& 0.168$\pm$0.015& \nodata& 0.312$\pm$0.018& \nodata& 0.36$\pm$ 0.06&  30&  0.35\\
U12894/A & +018 -004& 1.254$\pm$0.047& 0.225$\pm$0.012& 3.169$\pm$0.067& \nodata& \nodata& 2.779$\pm$0.103& 0.068$\pm$0.004& 0.028$\pm$0.002& 0.125$\pm$0.005& 0.038$\pm$0.002& 0.34$\pm$ 0.04& 127&  1.14\\
U290/A   & -006 +006& 2.238$\pm$0.202&   \nodata      & 1.493$\pm$0.107& \nodata& \nodata& 2.828$\pm$0.250& 0.145$\pm$0.023& \nodata& 0.326$\pm$0.029& \nodata& 0.40$\pm$ 0.09&  10&  0.16\\
U290/A   & +006 -006& 2.141$\pm$0.094& 0.118$\pm$0.022& 3.057$\pm$0.081& \nodata& \nodata& 2.832$\pm$0.125& 0.138$\pm$0.008& 0.029$\pm$0.004& 0.298$\pm$0.012& 0.047$\pm$0.005& 0.33$\pm$ 0.05&  17&  0.64\\
U685/A   & +000 +001& 2.446$\pm$0.103& 0.513$\pm$0.029& 6.048$\pm$0.144& \nodata& \nodata& 2.809$\pm$0.124& 0.169$\pm$0.009& 0.024$\pm$0.004& 0.547$\pm$0.019& 0.082$\pm$0.006& 0.11$\pm$ 0.05&  12&  1.34\\
U685/A   & +006 -011& 2.028$\pm$0.067& 0.467$\pm$0.017& 7.228$\pm$0.132& \nodata& \nodata& 2.748$\pm$0.097& 0.129$\pm$0.005& 0.031$\pm$0.003& 0.381$\pm$0.011& 0.108$\pm$0.005& 0.03$\pm$ 0.04&  67&  3.14\\
U685/B   & +006 -011& 2.057$\pm$0.073& 0.460$\pm$0.019& 7.218$\pm$0.139& \nodata& \nodata& 2.773$\pm$0.103& 0.146$\pm$0.006& 0.029$\pm$0.003& 0.377$\pm$0.011& 0.098$\pm$0.005& 0.07$\pm$ 0.04&  59&  3.35\\
U685/B   & +012 -004& 2.663$\pm$0.084& 0.128$\pm$0.006& 2.474$\pm$0.044& \nodata& \nodata& 2.601$\pm$0.089& 0.194$\pm$0.005& 0.024$\pm$0.002& 0.438$\pm$0.011& 0.066$\pm$0.003& 0.10$\pm$ 0.04&  77&  7.94\\
U1104/B  & +002 +007& 2.782$\pm$0.149& 0.409$\pm$0.024& 4.567$\pm$0.151& \nodata& 0.031$\pm$0.003& 2.788$\pm$0.157& 0.093$\pm$0.005& 0.018$\pm$0.002& 0.321$\pm$0.014& 0.072$\pm$0.005& 0.22$\pm$ 0.06&  24& 2.64\\
U1104/B  & +004 -001& 3.181$\pm$0.170& 0.245$\pm$0.016& 3.785$\pm$0.126& 0.051$\pm$0.004& 0.025$\pm$0.003& 2.868$\pm$0.161& 0.126$\pm$0.007& 0.029$\pm$0.003& 0.451$\pm$0.019& \nodata& 0.24$\pm$ 0.06&  20& 2.82\\
U1104/A  & +004 -001& 3.452$\pm$0.191& 0.508$\pm$0.031& 3.497$\pm$0.120& \nodata& \nodata& 2.932$\pm$0.114& 0.176$\pm$0.011& 0.033$\pm$0.005& 0.548$\pm$0.025& 0.074$\pm$0.007& 0.32$\pm$ 0.06&  13&  2.05\\
U1175/A  & -004 +007& 2.751$\pm$0.203&   \nodata      & 1.024$\pm$0.080& \nodata& \nodata& 2.823$\pm$0.217& 0.252$\pm$0.028& \nodata& 0.454$\pm$0.036& \nodata& 0.09$\pm$ 0.08&   7&  0.21\\
U1175/A  & -002 -001& 2.608$\pm$0.152&   \nodata      & 1.561$\pm$0.071& \nodata& \nodata& 2.824$\pm$0.169& 0.155$\pm$0.013& 0.088$\pm$0.010& 0.419$\pm$0.022& \nodata& 0.18$\pm$ 0.06&  16&  0.34\\
U1175/A  & +000 -007& 2.139$\pm$0.128& 0.297$\pm$0.039& 2.351$\pm$0.093& \nodata& \nodata& 2.796$\pm$0.172& 0.133$\pm$0.015& 0.027$\pm$0.010& 0.353$\pm$0.022& \nodata& 0.03$\pm$ 0.06&  21&  0.40\\
U1281/A  & +008 +013& 2.036$\pm$0.170&    \nodata     & 2.764$\pm$0.135& \nodata& \nodata& 2.795$\pm$0.231& 0.234$\pm$0.024& \nodata& 0.420$\pm$0.032& 0.058$\pm$0.013& 0.18$\pm$ 0.08&  11&  0.48\\
HKK L14/A& +002 -004& 1.232$\pm$0.071& 0.156$\pm$0.016& 3.289$\pm$0.114& \nodata& \nodata& 3.068$\pm$0.176& 0.124$\pm$0.007& 0.023$\pm$0.003& 0.239$\pm$0.011& 0.056$\pm$0.005& 0.00$\pm$ 0.06&  93&  1.66\\
U2023/A  & -036 -010& 2.620$\pm$0.173&   \nodata      & 3.092$\pm$0.123& \nodata& \nodata& 2.727$\pm$0.176& 0.308$\pm$0.020& \nodata& 0.433$\pm$0.025& 0.066$\pm$0.009& 0.30$\pm$ 0.06&  86&  1.34\\
U2023/C  & -029 +023& 4.156$\pm$0.716&   \nodata      & 1.790$\pm$0.247& \nodata& \nodata& 2.820$\pm$0.516& 0.387$\pm$0.073& \nodata& 0.538$\pm$0.085& \nodata& 0.33$\pm$ 0.18&  43&  0.18\\
U2023/B  & -028 +023& 4.448$\pm$0.594&   \nodata      & 1.473$\pm$0.174& \nodata& \nodata& 2.828$\pm$0.400& 0.315$\pm$0.054& \nodata& 0.611$\pm$0.071& \nodata& 0.37$\pm$ 0.14&  50&  0.23\\
U2023/C  & -020 +023& 3.871$\pm$0.276&   \nodata      & 1.921$\pm$0.097& \nodata& \nodata& 2.855$\pm$0.212& 0.337$\pm$0.025& \nodata& 0.755$\pm$0.044& 0.057$\pm$0.011& 0.16$\pm$ 0.07&  60&  0.55\\
U2023/B  & -014 +024& 2.729$\pm$0.094& 0.157$\pm$0.010& 3.345$\pm$0.065& 0.025$\pm$0.002& 0.012$\pm$0.002& 2.808$\pm$0.103& 0.250$\pm$0.008& 0.029$\pm$0.002& 0.439$\pm$0.013& 0.080$\pm$0.004& 0.11$\pm$ 0.04&  84&  2.76\\
U2023/C  & -013 +022& 3.227$\pm$0.191& 0.315$\pm$0.027& 4.392$\pm$0.160& 0.060$\pm$0.006& 0.013$\pm$0.005& 2.882$\pm$0.179& 0.294$\pm$0.016& \nodata& 0.701$\pm$0.033& 0.080$\pm$0.007& 0.12$\pm$ 0.06&  84& 1.28\\
U2023/B  & -008 +025& 2.873$\pm$0.088& 0.246$\pm$0.008& 3.977$\pm$0.069& 0.045$\pm$0.002& 0.019$\pm$0.001& 2.808$\pm$0.094& 0.229$\pm$0.006& 0.027$\pm$0.001& 0.473$\pm$0.012& 0.083$\pm$0.003& 0.19$\pm$ 0.04& 174& 12.09\\
U2023/C  & -006 +022& 1.968$\pm$0.101& 0.288$\pm$0.015& 5.325$\pm$0.171& 0.024$\pm$0.001& 0.019$\pm$0.001& 2.831$\pm$0.154& 0.163$\pm$0.008& 0.029$\pm$0.002& 0.265$\pm$0.011& 0.096$\pm$0.006& 0.28$\pm$ 0.05&  84&  6.87\\
U2023/C  & +028 +020& 3.784$\pm$0.249&   \nodata      & 2.115$\pm$0.091& \nodata& \nodata& 2.828$\pm$0.196& 0.331$\pm$0.022& \nodata& 0.868$\pm$0.047& \nodata& 0.16$\pm$ 0.07&  77&  0.95\\
U3647/A  & -006 -018& 1.719$\pm$0.069& 0.288$\pm$0.014& 5.937$\pm$0.113& 0.019$\pm$0.002& 0.014$\pm$0.002& 2.731$\pm$0.134& 0.168$\pm$0.007& 0.032$\pm$0.002& 0.242$\pm$0.009& 0.089$\pm$0.005& 0.34$\pm$ 0.05& 314&  5.19\\
U3647/A  & +004 +002& 2.416$\pm$0.123& 0.325$\pm$0.027& 5.940$\pm$0.150& 0.032$\pm$0.005& 0.026$\pm$0.005& 2.867$\pm$0.164& 0.255$\pm$0.013& 0.017$\pm$0.011& 0.362$\pm$0.016& 0.098$\pm$0.008& 0.27$\pm$ 0.06&  78&  1.80\\
U3647/A  & +006 +006& 1.793$\pm$0.080& 0.260$\pm$0.017& 5.568$\pm$0.119& \nodata& \nodata& 2.712$\pm$0.141& 0.199$\pm$0.009& 0.030$\pm$0.003& 0.349$\pm$0.014& 0.082$\pm$0.006& 0.18$\pm$ 0.05&  74&  2.75\\
U3672/A  & -052 +091& 2.461$\pm$0.266&   \nodata      & 1.172$\pm$0.121& \nodata& \nodata& 2.899$\pm$0.314& \nodata& \nodata& 0.275$\pm$0.037& \nodata& 0.40$\pm$ 0.10&  27&  0.17\\
U4117/A  & -011 +017& 2.119$\pm$0.149&   \nodata      & 3.828$\pm$0.163& \nodata& \nodata& 2.872$\pm$0.206& 0.120$\pm$0.020& \nodata& 0.352$\pm$0.027& \nodata& 0.15$\pm$ 0.07&  26&  0.47\\
U4117/B  & -001 +008& 3.190$\pm$0.226&   \nodata      & 3.138$\pm$0.140& \nodata& \nodata& 2.827$\pm$0.208& 0.163$\pm$0.020& \nodata& 0.401$\pm$0.029& \nodata& 0.14$\pm$ 0.07&  13&  0.31\\
U4117/B  & +012 +003& 1.099$\pm$0.069& 0.317$\pm$0.025& 5.175$\pm$0.184& \nodata& \nodata& 2.770$\pm$0.169& 0.057$\pm$0.010& \nodata& 0.172$\pm$0.013& 0.054$\pm$0.008& 0.05$\pm$ 0.06&  59&  0.64\\
U4483/A  & -001 +016& 0.779$\pm$0.041& 0.261$\pm$0.013& 4.316$\pm$0.138& \nodata& 0.012$\pm$0.001& 2.773$\pm$0.151& 0.034$\pm$0.002& 0.027$\pm$0.002& 0.088$\pm$0.004& 0.038$\pm$0.002& 0.12$\pm$ 0.05& 161& 18.42\\
U4483/A  & +001 -006& 1.183$\pm$0.075& 0.248$\pm$0.023& 3.943$\pm$0.142& \nodata& \nodata& 2.826$\pm$0.173& 0.052$\pm$0.006& 0.022$\pm$0.005& 0.134$\pm$0.009& 0.044$\pm$0.005& 0.09$\pm$ 0.06&  63&  2.56\\
CGCG 007-025/A& -008 +007& 1.207$\pm$0.050& 0.340$\pm$0.015& 5.800$\pm$0.132& 0.025$\pm$0.002& 0.022$\pm$0.002& 2.807$\pm$0.128& 0.065$\pm$0.003& 0.026$\pm$0.002& 0.194$\pm$0.007& 0.041$\pm$0.002& 0.22$\pm$ 0.05& 179&  6.19\\
CGCG 007-025/A& -004 +000& 0.949$\pm$0.036& 0.404$\pm$0.015& 6.792$\pm$0.147& 0.024$\pm$0.001& 0.015$\pm$0.001& 2.759$\pm$0.122& 0.055$\pm$0.002& 0.027$\pm$0.001& 0.141$\pm$0.005& 0.050$\pm$0.003& 0.31$\pm$ 0.04& 230& 36.38\\
CGCG 007-025/A& +000 -009& 1.242$\pm$0.050& 0.369$\pm$0.016& 5.340$\pm$0.120& 0.028$\pm$0.002& 0.019$\pm$0.001& 2.894$\pm$0.131& 0.074$\pm$0.004& 0.027$\pm$0.002& 0.171$\pm$0.006& 0.040$\pm$0.002& 0.12$\pm$ 0.05&  68&  7.56\\
U5288/A  & -007 +010& 2.130$\pm$0.111& 0.407$\pm$0.021& 5.240$\pm$0.168& 0.044$\pm$0.003& 0.020$\pm$0.002& 2.817$\pm$0.154& 0.204$\pm$0.009& 0.027$\pm$0.002& 0.531$\pm$0.022& 0.089$\pm$0.006& 0.17$\pm$ 0.05&  84&  6.87\\
U5288/A  & -004 +006& 2.921$\pm$0.153& 0.395$\pm$0.021& 4.769$\pm$0.155& 0.060$\pm$0.004& 0.017$\pm$0.003& 2.819$\pm$0.156& 0.256$\pm$0.011& 0.022$\pm$0.003& 0.636$\pm$0.027& 0.083$\pm$0.006& 0.21$\pm$ 0.05&  32&  6.05\\
U5288/A  & -000 +000& 3.410$\pm$0.187& 0.223$\pm$0.017& 3.956$\pm$0.136& \nodata& \nodata& 2.813$\pm$0.162& 0.310$\pm$0.016& 0.021$\pm$0.005& 0.669$\pm$0.030& 0.090$\pm$0.008& 0.16$\pm$ 0.06&  18&  2.06\\
U5288/B  & +006 -013& 3.940$\pm$0.179& 0.633$\pm$0.038& 4.344$\pm$0.119& 0.096$\pm$0.010& 0.037$\pm$0.010& 2.750$\pm$0.131& 0.280$\pm$0.018& \nodata& 0.842$\pm$0.033& 0.096$\pm$0.010& 0.16$\pm$ 0.05&  11& 2.34\\
U5288/A  & +023 -034& 2.577$\pm$0.191& 0.276$\pm$0.043& 5.333$\pm$0.231& \nodata& \nodata& 2.828$\pm$0.215& 0.085$\pm$0.020& \nodata& 0.296$\pm$0.028& 0.102$\pm$0.017& 0.15$\pm$ 0.07& -99&  0.60\\
U5288/C  & +026 -033& 2.331$\pm$0.208&   \nodata      & 4.973$\pm$0.267& \nodata& \nodata& 2.776$\pm$0.252& 0.128$\pm$0.022& \nodata& 0.312$\pm$0.032& \nodata& 0.24$\pm$ 0.09& 107&  0.35\\
UA292/C  & +008 +012& 0.913$\pm$0.036& 0.121$\pm$0.005& 1.771$\pm$0.042& \nodata& \nodata& 2.790$\pm$0.110& 0.055$\pm$0.002& 0.026$\pm$0.002& 0.103$\pm$0.004& 0.018$\pm$0.001& 0.07$\pm$ 0.04& 124 & 2.37\\
UA292/C  & +026 +004& 1.162$\pm$0.052& \nodata & 0.416$\pm$0.021& \nodata& \nodata& 2.781$\pm$0.123& 0.093$\pm$0.009& 0.017$\pm$0.006& 0.178$\pm$0.011& \nodata& 0.02$\pm$ 0.05&  64& 0.65\\
UA292/D  & +035 -003& 0.553$\pm$0.022& 0.197$\pm$0.009& 2.480$\pm$0.059& \nodata& \nodata& 2.776$\pm$0.110& 0.038$\pm$0.002& 0.025$\pm$0.002& 0.059$\pm$0.003& 0.019$\pm$0.001& 0.12$\pm$ 0.04& 121 & 3.90\\
UA292/C  & +038 +000& 0.849$\pm$0.038& \nodata & 1.642$\pm$0.044& \nodata& \nodata& 2.705$\pm$0.117& 0.059$\pm$0.009& 0.020$\pm$0.006& 0.127$\pm$0.010& \nodata& 0.02$\pm$ 0.05& 114& 1.01\\
U8651/A  & +045 +036& 2.005$\pm$0.078& 0.166$\pm$0.008& 2.730$\pm$0.065& 0.007$\pm$0.003& 0.022$\pm$0.003& 2.839$\pm$0.113& 0.092$\pm$0.005& 0.023$\pm$0.003& 0.176$\pm$0.006& 0.044$\pm$0.003& 0.03$\pm$ 0.04&  75&  8.21\\
U9240/A  & +004 -037& 1.603$\pm$0.069& 0.393$\pm$0.017& 5.693$\pm$0.154& 0.019$\pm$0.001& 0.017$\pm$0.001& 2.695$\pm$0.115& 0.074$\pm$0.003& 0.027$\pm$0.002& 0.175$\pm$0.006& 0.074$\pm$0.004& 0.00$\pm$ 0.04& 167&  9.03\\
U9992/A  & +007 +010& 2.676$\pm$0.262& \nodata & 0.865$\pm$0.105& \nodata& \nodata& 2.745$\pm$0.281& 0.280$\pm$0.037& \nodata& 0.749$\pm$0.065& \nodata& 0.00$\pm$ 0.10&  47 & 0.24\\
U9992/A  & +011 -003& 1.943$\pm$0.094& 0.177$\pm$0.022& 3.756$\pm$0.111& \nodata& \nodata& 2.805$\pm$0.137& 0.199$\pm$0.012& 0.026$\pm$0.006& 0.578$\pm$0.024& 0.056$\pm$0.007& 0.13$\pm$ 0.05&  47 & 0.82\\
U9992/A  & +013 -012& 3.518$\pm$0.155& 0.079$\pm$0.017& 1.611$\pm$0.048& \nodata& \nodata& 2.830$\pm$0.129& 0.293$\pm$0.013& 0.020$\pm$0.005& 0.767$\pm$0.027& 0.055$\pm$0.006& 0.15$\pm$ 0.05&  35 & 1.36\\
U10445/B & -034 +053& 3.300$\pm$0.399&   \nodata      & 5.395$\pm$0.414& \nodata& \nodata& 2.860$\pm$0.367& 0.407$\pm$0.069& \nodata& 0.721$\pm$0.088& \nodata& 0.06$\pm$ 0.12& 107&  0.25\\
U10445/A & -026 -002& 3.001$\pm$0.132& 0.176$\pm$0.010& 2.722$\pm$0.077& 0.055$\pm$0.003& 0.017$\pm$0.002& 2.793$\pm$0.122& 0.374$\pm$0.013& 0.027$\pm$0.002& 0.623$\pm$0.021& 0.075$\pm$0.004& 0.05$\pm$ 0.05&  81&  5.66\\
U10445/B & -025 +050& 3.277$\pm$0.255&   \nodata      & 1.672$\pm$0.106& \nodata& \nodata& 2.482$\pm$0.202& 0.353$\pm$0.036& \nodata& 0.844$\pm$0.058& \nodata& 0.00$\pm$ 0.08&  31&  0.51\\
U10445/B & -019 +048& 3.462$\pm$0.148& 0.221$\pm$0.011& 2.865$\pm$0.080& 0.106$\pm$0.005& 0.015$\pm$0.002& 2.785$\pm$0.117& 0.363$\pm$0.012& 0.026$\pm$0.002& 0.782$\pm$0.024& 0.065$\pm$0.004& 0.04$\pm$ 0.04&  98&  5.64\\
U10445/B & -015 +047& 3.173$\pm$0.142& 0.198$\pm$0.013& 2.843$\pm$0.082& 0.077$\pm$0.005& 0.011$\pm$0.003& 2.802$\pm$0.126& 0.390$\pm$0.014& 0.032$\pm$0.003& 0.771$\pm$0.027& 0.068$\pm$0.005& 0.02$\pm$ 0.05&  95&  2.98\\
U10445/B & -010 +045& 3.714$\pm$0.229&   \nodata      & 2.811$\pm$0.116& \nodata& \nodata& 2.811$\pm$0.179& 0.347$\pm$0.022& \nodata& 0.522$\pm$0.029& 0.062$\pm$0.010& 0.22$\pm$ 0.06&  96&  8.97\\
U10445/A & -003 +010& 3.349$\pm$0.273&   \nodata      & 2.486$\pm$0.142& 0.124$\pm$0.021& \nodata& 2.840$\pm$0.241& 0.538$\pm$0.045& \nodata& 0.414$\pm$0.037& \nodata& 0.14$\pm$ 0.08&  20&  0.63\\
U11755/A & -003 +003& 2.067$\pm$0.067& 0.246$\pm$0.008& 4.307$\pm$0.082& 0.023$\pm$0.002& 0.014$\pm$0.001& 2.809$\pm$0.096& 0.310$\pm$0.009& 0.031$\pm$0.002& 0.385$\pm$0.010& 0.103$\pm$0.004& 0.22$\pm$ 0.04&  45& 21.56\\
U11755/A & -001 +000& 2.938$\pm$0.099& 0.301$\pm$0.012& 3.287$\pm$0.065& 0.051$\pm$0.003& 0.013$\pm$0.003& 2.821$\pm$0.100& 0.487$\pm$0.015& 0.031$\pm$0.003& 0.784$\pm$0.022& 0.084$\pm$0.004& 0.23$\pm$ 0.04&  12&  7.33\\
U12713/B & -009 -003& 1.644$\pm$0.090&   \nodata      & 3.403$\pm$0.092& \nodata& \nodata& 2.671$\pm$0.156& 0.073$\pm$0.009& 0.034$\pm$0.007& 0.267$\pm$0.015& 0.043$\pm$0.007& 0.22$\pm$ 0.06&  22&  1.76\\
U12713/A & -009 -003& 1.623$\pm$0.084&   \nodata      & 3.552$\pm$0.093& \nodata& \nodata& 2.790$\pm$0.160& 0.092$\pm$0.015& 0.022$\pm$0.011& 0.249$\pm$0.019& 0.036$\pm$0.011& 0.18$\pm$ 0.06&  25&  0.91\\
U12713/A & -001 -003& 2.122$\pm$0.083& 0.197$\pm$0.012& 3.277$\pm$0.064& \nodata& \nodata& 2.803$\pm$0.138& 0.098$\pm$0.007& 0.026$\pm$0.004& 0.300$\pm$0.012& 0.053$\pm$0.005& 0.25$\pm$ 0.05&  42&  2.18\\
U12713/B & +004 +004& 2.163$\pm$0.088&   \nodata      & 2.128$\pm$0.045& \nodata& \nodata& 2.711$\pm$0.136& 0.117$\pm$0.007& 0.019$\pm$0.004& 0.404$\pm$0.016& \nodata& 0.17$\pm$ 0.05&  29&  4.34\\
\enddata
\tablecomments{EW of H$\beta$ is listed in units of \AA.  H$\beta$ flux is the observed quantity 
in units of erg s$^{-1}$ cm$^{-2}$; no corrections have been applied for
extinction and some observations were obtained in non-photometric conditions.}
\end{deluxetable}

\begin{deluxetable}{lrcccccc}
\tabcolsep 2pt
\tablewidth{0pt}
\tabletypesize{\small}
\tablecaption{HII Region Line Ratios\label{tab:ratio}}
\tablehead{
\colhead{Galaxy/} & \colhead{Location} &\colhead{[SII]}& \colhead{[OIII]} & \colhead{Log}& \colhead{Log} & \colhead{Log} & \colhead{Log}  \\
 \colhead{Slit }& \colhead{EW  NS } & \colhead{ratio} & \colhead{ratio}& \colhead{[OII]+[OIII]}& \colhead{[OIII]/[OII]} &\colhead{[NII]/[OII]}& \colhead{[SII]/[OII]}
}
\startdata
U12894/A & -006 -005& 1.53$\pm$0.26&    \nodata       & 0.705$\pm$0.013&  0.356$\pm$0.031& -1.287$\pm$0.076& -0.930$\pm$0.045\\
U12894/A & +004 -004& 1.28$\pm$0.15&    \nodata       & 0.689$\pm$0.013&  0.217$\pm$0.029& -1.041$\pm$0.046& -0.772$\pm$0.035\\
U12894/A & +018 -004& 1.40$\pm$0.11&  60.9 $\pm$  9.4 & 0.646$\pm$0.008&  0.403$\pm$0.019& -1.266$\pm$0.030& -1.001$\pm$0.024\\
U290/A   & -006 +006& 1.23$\pm$0.22&    \nodata       & 0.577$\pm$0.026& -0.184$\pm$0.049& -1.188$\pm$0.079& -0.837$\pm$0.055\\
U290/A   & +006 -006& 1.55$\pm$0.13&    \nodata       & 0.716$\pm$0.010&  0.155$\pm$0.022& -1.191$\pm$0.032& -0.856$\pm$0.026\\
U685/A   & +000 +001& 1.36$\pm$0.10&    \nodata       & 0.929$\pm$0.009&  0.393$\pm$0.021& -1.161$\pm$0.030& -0.650$\pm$0.024\\
U685/A   & +006 -011& 1.35$\pm$0.08&  81.2 $\pm$  6.6 & 0.966$\pm$0.007&  0.552$\pm$0.016& -1.196$\pm$0.022& -0.726$\pm$0.019\\
U685/B   & +006 -011& 1.45$\pm$0.09&  85.9 $\pm$ 10.4 & 0.967$\pm$0.007&  0.545$\pm$0.018& -1.149$\pm$0.024& -0.737$\pm$0.020\\
U685/B   & +012 -004& 1.39$\pm$0.07&  88.4 $\pm$ 12.7 & 0.711$\pm$0.008& -0.032$\pm$0.016& -1.138$\pm$0.018& -0.784$\pm$0.018\\
U1104/B  & +002 +007& 1.45$\pm$0.12&  86.2 $\pm$ 14.9 & 0.866$\pm$0.013&  0.215$\pm$0.027& -1.476$\pm$0.033& -0.938$\pm$0.030\\
U1104/B  & +004 -001& 1.45$\pm$0.12&  84.1 $\pm$ 17.0 & 0.843$\pm$0.013&  0.076$\pm$0.027& -1.402$\pm$0.034& -0.848$\pm$0.030\\
U1104/A  & +004 -001& 1.36$\pm$0.12&    \nodata       & 0.842$\pm$0.014&  0.006$\pm$0.028& -1.293$\pm$0.036& -0.799$\pm$0.021\\
U1175/A  & -004 +007& 1.82$\pm$0.30&    \nodata       & 0.577$\pm$0.025& -0.429$\pm$0.047& -1.038$\pm$0.058& -0.782$\pm$0.047\\
U1175/A  & -002 -001& 1.31$\pm$0.14&    \nodata       & 0.620$\pm$0.017& -0.223$\pm$0.032& -1.226$\pm$0.044& -0.794$\pm$0.034\\
U1175/A  & +000 -007& 1.89$\pm$0.25&    \nodata       & 0.652$\pm$0.015&  0.041$\pm$0.031& -1.206$\pm$0.056& -0.782$\pm$0.038\\
U1281/A  & +008 +013& 1.47$\pm$0.23&    \nodata       & 0.681$\pm$0.020&  0.133$\pm$0.042& -0.940$\pm$0.058& -0.686$\pm$0.049\\
HKK L14/A& +002 -004& 1.32$\pm$0.13&    \nodata       & 0.655$\pm$0.013&  0.426$\pm$0.029& -0.997$\pm$0.035& -0.712$\pm$0.032\\
U2023/A  & -036 -010& 1.27$\pm$0.15&    \nodata       & 0.757$\pm$0.016&  0.072$\pm$0.034& -0.930$\pm$0.040& -0.782$\pm$0.038\\
U2023/C  & -029 +023& 1.30$\pm$0.41&    \nodata       & 0.774$\pm$0.055& -0.366$\pm$0.096& -1.031$\pm$0.111& -0.888$\pm$0.102\\
U2023/B  & -028 +023& 0.97$\pm$0.23&    \nodata       & 0.772$\pm$0.045& -0.480$\pm$0.077& -1.150$\pm$0.094& -0.862$\pm$0.077\\
U2023/C  & -020 +023& 1.32$\pm$0.15&    \nodata       & 0.763$\pm$0.022& -0.304$\pm$0.038& -1.060$\pm$0.045& -0.710$\pm$0.040\\
U2023/B  & -014 +024& 1.40$\pm$0.08&    \nodata       & 0.783$\pm$0.008&  0.088$\pm$0.017& -1.038$\pm$0.020& -0.794$\pm$0.020\\
U2023/C  & -013 +022& 1.46$\pm$0.14&    \nodata       & 0.882$\pm$0.014&  0.134$\pm$0.030& -1.040$\pm$0.035& -0.663$\pm$0.033\\
U2023/B  & -008 +025& 1.35$\pm$0.07& 102.0 $\pm$  8.0 & 0.836$\pm$0.007&  0.141$\pm$0.015& -1.099$\pm$0.018& -0.783$\pm$0.017\\
U2023/C  & -006 +022& 1.35$\pm$0.11& 106.5 $\pm$  7.2 & 0.863$\pm$0.012&  0.432$\pm$0.026& -1.082$\pm$0.031& -0.871$\pm$0.029\\
U2023/C  & +028 +020& 1.33$\pm$0.14&    \nodata       & 0.771$\pm$0.020& -0.253$\pm$0.034& -1.058$\pm$0.041& -0.639$\pm$0.037\\
U3647/A  & -006 -018& 1.44$\pm$0.11& 116.4 $\pm$ 16.1 & 0.884$\pm$0.008&  0.538$\pm$0.019& -1.010$\pm$0.025& -0.851$\pm$0.024\\
U3647/A  & +004 +002& 1.37$\pm$0.12&    \nodata       & 0.922$\pm$0.010&  0.391$\pm$0.025& -0.977$\pm$0.031& -0.824$\pm$0.029\\
U3647/A  & +006 +006& 1.48$\pm$0.12&    \nodata       & 0.867$\pm$0.008&  0.492$\pm$0.022& -0.955$\pm$0.028& -0.711$\pm$0.026\\
U3672/A  & -052 +091& 1.81$\pm$0.52&    \nodata       & 0.560$\pm$0.035& -0.322$\pm$0.065&       \nodata   & -0.952$\pm$0.075\\
U4117/A  & -011 +017& 1.73$\pm$0.28&    \nodata       & 0.774$\pm$0.016&  0.257$\pm$0.036& -1.247$\pm$0.079& -0.780$\pm$0.045\\
U4117/B  & -001 +008& 1.55$\pm$0.22&    \nodata       & 0.801$\pm$0.018& -0.007$\pm$0.036& -1.292$\pm$0.062& -0.901$\pm$0.044\\
U4117/B  & +012 +003& 1.05$\pm$0.16&    \nodata       & 0.798$\pm$0.014&  0.673$\pm$0.031& -1.285$\pm$0.081& -0.805$\pm$0.043\\
U4483/A  & -001 +016& 1.38$\pm$0.11&  56.0 $\pm$  3.4 & 0.707$\pm$0.012&  0.744$\pm$0.027& -1.360$\pm$0.034& -0.947$\pm$0.030\\
U4483/A  & +001 -006& 1.31$\pm$0.18&    \nodata       & 0.710$\pm$0.014&  0.523$\pm$0.032& -1.357$\pm$0.057& -0.946$\pm$0.040\\
CGCG 007-025/A& -008 +007& 1.40$\pm$0.10&  71.6 $\pm$  5.6 & 0.846$\pm$0.009&  0.682$\pm$0.021& -1.269$\pm$0.027& -0.794$\pm$0.024\\
CGCG 007-025/A& -004 +000& 1.31$\pm$0.09&  60.1 $\pm$  2.5 & 0.889$\pm$0.008&  0.855$\pm$0.019& -1.237$\pm$0.023& -0.828$\pm$0.023\\
CGCG 007-025/A& +000 -009& 1.34$\pm$0.10&  82.1 $\pm$  6.6 & 0.818$\pm$0.009&  0.633$\pm$0.020& -1.225$\pm$0.029& -0.861$\pm$0.023\\
U5288/A  & -007 +010& 1.37$\pm$0.11&  77.0 $\pm$  6.2 & 0.867$\pm$0.012&  0.391$\pm$0.027& -1.019$\pm$0.030& -0.603$\pm$0.029\\
U5288/A  & -004 +006& 1.50$\pm$0.12&  73.4 $\pm$  8.2 & 0.886$\pm$0.012&  0.213$\pm$0.027& -1.057$\pm$0.029& -0.662$\pm$0.029\\
U5288/A  & -000 +000& 1.35$\pm$0.12&  77.6 $\pm$ 18.4 & 0.867$\pm$0.014&  0.065$\pm$0.028& -1.041$\pm$0.033& -0.707$\pm$0.031\\
U5288/B  & +006 -013& 1.49$\pm$0.12&    \nodata       & 0.918$\pm$0.011&  0.042$\pm$0.023& -1.148$\pm$0.034& -0.670$\pm$0.026\\
U5288/A  & +023 -034& 1.53$\pm$0.30&    \nodata       & 0.898$\pm$0.016&  0.316$\pm$0.037& -1.482$\pm$0.107& -0.940$\pm$0.052\\
U5288/C  & +026 -033& 0.87$\pm$0.18&    \nodata       & 0.864$\pm$0.020&  0.329$\pm$0.045& -1.260$\pm$0.084& -0.873$\pm$0.059\\
UA292/C  & +004 +013& 1.36$\pm$0.09&  54.4 $\pm$  3.2 & 0.438$\pm$0.009&  0.444$\pm$0.020& -1.144$\pm$0.024& -0.931$\pm$0.023\\
UA292/C  & +008 +012& 1.58$\pm$0.11&  61.1 $\pm$  6.5 & 0.429$\pm$0.009&  0.288$\pm$0.020& -1.220$\pm$0.023& -0.948$\pm$0.024\\
UA292/C  & +026 +004& 1.44$\pm$0.18&    \nodata       & 0.198$\pm$0.015& -0.446$\pm$0.030& -1.097$\pm$0.046& -0.815$\pm$0.033\\
UA292/D  & +035 -003& 1.57$\pm$0.16&  51.7 $\pm$  4.5 & 0.482$\pm$0.009&  0.652$\pm$0.020& -1.163$\pm$0.029& -0.972$\pm$0.028\\
UA292/C  & +038 +000& 1.59$\pm$0.27&    \nodata       & 0.396$\pm$0.010&  0.286$\pm$0.023& -1.158$\pm$0.069& -0.825$\pm$0.039\\
U8651/A  & +045 +036& 1.38$\pm$0.10& 109.2 $\pm$ 17.7 & 0.675$\pm$0.009&  0.134$\pm$0.020& -1.338$\pm$0.029& -1.057$\pm$0.022\\
U9240/A  & +004 -037& 1.43$\pm$0.09&  90.4 $\pm$  6.2 & 0.863$\pm$0.010&  0.550$\pm$0.022& -1.336$\pm$0.026& -0.962$\pm$0.024\\
U9992/A  & +007 +010& 1.81$\pm$0.31&    \nodata       & 0.549$\pm$0.035& -0.490$\pm$0.068& -0.980$\pm$0.072& -0.553$\pm$0.057\\
U9992/A  & +011 -003& 1.44$\pm$0.12&  98.8 $\pm$ 52.1 & 0.756$\pm$0.011&  0.286$\pm$0.025& -0.990$\pm$0.034& -0.526$\pm$0.028\\
U9992/A  & +013 -012& 1.49$\pm$0.11&    \nodata       & 0.710$\pm$0.014& -0.339$\pm$0.027& -1.079$\pm$0.027& -0.662$\pm$0.025\\
U10445/B & -034 +053& 1.33$\pm$0.33&    \nodata       & 0.939$\pm$0.029&  0.213$\pm$0.062& -0.909$\pm$0.091& -0.661$\pm$0.075\\
U10445/A & -026 -002& 1.46$\pm$0.10& 104.7 $\pm$ 24.3 & 0.758$\pm$0.012& -0.042$\pm$0.023& -0.904$\pm$0.024& -0.683$\pm$0.024\\
U10445/B & -025 +050& 1.44$\pm$0.20&    \nodata       & 0.695$\pm$0.024& -0.292$\pm$0.044& -0.968$\pm$0.056& -0.589$\pm$0.045\\
U10445/B & -019 +048& 1.40$\pm$0.09&    \nodata       & 0.801$\pm$0.012& -0.082$\pm$0.022& -0.979$\pm$0.023& -0.646$\pm$0.023\\
U10445/B & -015 +047& 1.49$\pm$0.10& 101.5 $\pm$ 32.8 & 0.779$\pm$0.012& -0.048$\pm$0.023& -0.910$\pm$0.025& -0.614$\pm$0.025\\
U10445/B & -010 +045& 1.50$\pm$0.16&    \nodata       & 0.815$\pm$0.017& -0.121$\pm$0.032& -1.030$\pm$0.038& -0.852$\pm$0.036\\
U10445/A & -003 +010& 1.29$\pm$0.23&    \nodata       & 0.766$\pm$0.023& -0.129$\pm$0.043& -0.794$\pm$0.051& -0.908$\pm$0.053\\
U11755/A & -003 +003& 1.48$\pm$0.08& 134.6 $\pm$ 8.8  & 0.804$\pm$0.007&  0.319$\pm$0.016& -0.824$\pm$0.019& -0.730$\pm$0.018\\
U11755/A & -001 +000& 1.48$\pm$0.08& 109.6 $\pm$ 25.7 & 0.794$\pm$0.008&  0.049$\pm$0.017& -0.781$\pm$0.020& -0.574$\pm$0.019\\
U12713/B & -009 -003& 1.50$\pm$0.17&    \nodata       & 0.703$\pm$0.011&  0.316$\pm$0.027& -1.353$\pm$0.059& -0.789$\pm$0.034\\
U12713/A & -009 -003& 2.00$\pm$0.34&    \nodata       & 0.714$\pm$0.011&  0.340$\pm$0.025& -1.247$\pm$0.074& -0.814$\pm$0.040\\
U12713/A & -001 -003& 1.46$\pm$0.12&  84.0 $\pm$ 19.4 & 0.732$\pm$0.008&  0.189$\pm$0.019& -1.336$\pm$0.035& -0.850$\pm$0.024\\
U12713/B & +004 +004& 1.40$\pm$0.11&    \nodata       & 0.633$\pm$0.010& -0.007$\pm$0.020& -1.267$\pm$0.031& -0.729$\pm$0.025\\
\enddata
\tablecomments{[S II] ratio = I(6717)/I(6731); [O III] ratio = I(4959+5007)/I(4363).}
\end{deluxetable}

\begin{deluxetable}{lrlccccccc}
\tablewidth{0pt}
\tabcolsep 2pt
\tabletypesize{\scriptsize}
\tablecaption{HII Region Abundances\label{tab:abund}}
\tablehead{
\colhead{} & \colhead{} & \colhead{}&\colhead{12+}& \colhead{12+} & \colhead{}& \colhead{} & \colhead{} & \colhead{}  & \colhead{} \\
\colhead{Galaxy} & \colhead{Location} & \colhead{T$_{\rm e}$[OIII]}&\colhead{Log(O/H)}& \colhead{Log(O/H)} & \colhead{12 + Log O/H}& \colhead{Log N/O} & \colhead{Log Ne/O} & \colhead{Log S/O}  & \colhead{Log Ar/O} \\
 \colhead{}& \colhead{EW  NS } & \colhead{K}& \colhead{McGaugh} & \colhead{Pilyugin}& \colhead{}& \colhead{} &\colhead{}& \colhead{} & \colhead{}
}
\startdata
U12894 & -006 -005 & 14500 $\pm$ 2000           &  7.72&  7.65&  7.71$\pm$ 0.10& -1.52$\pm$ 0.14& -0.69$\pm$ 0.24&      \nodata   &    \nodata    \\
U12894 & +004 -004 & 13900 $\pm$ 2000           &  7.75&  7.71&  7.75$\pm$ 0.10& -1.37$\pm$ 0.15&  \nodata       &      \nodata   &    \nodata    \\
U12894 & +018 -004 & 15900 $\pm$ $^{1390}_{1020}$& 7.65&  7.54&  7.56$\pm$ 0.04& -1.54$\pm$ 0.08& -0.77$\pm$ 0.12&      \nodata   & -2.25$\pm$ 0.10\\
U290   & -006 +006 & 12500 $\pm$ 1500           &  7.75&  7.89&  7.75$\pm$ 0.10& -1.57$\pm$ 0.16&  \nodata       &      \nodata   &    \nodata    \\
U290   & +006 -006 & 13200 $\pm$ 1500           &  7.82&  7.79&  7.82$\pm$ 0.10& -1.45$\pm$ 0.11& -1.00$\pm$ 0.22&      \nodata   & -2.28$\pm$ 0.15\\
U685   & +000 +001 & 11500 $\pm$ 1500           &  8.18&  7.95&  8.18$\pm$ 0.10& -1.46$\pm$ 0.15& -0.89$\pm$ 0.28&      \nodata   & -2.23$\pm$ 0.18\\
U685   & +006 -011 & 13970 $\pm$ $^{500}_{430}$ &  8.25&  7.94&  8.00$\pm$ 0.03& -1.45$\pm$ 0.06& -0.78$\pm$ 0.07&      \nodata   & -2.08$\pm$ 0.08\\
U685   & +006 -011 & 13630 $\pm$ $^{790}_{610}$ &  8.25&  7.94&  8.02$\pm$ 0.04& -1.40$\pm$ 0.07& -0.79$\pm$ 0.10&      \nodata   & -2.12$\pm$ 0.09\\
U685   & +012 -004 & 13490 $\pm$ $^{930}_{690}$ &  7.90&  7.93&  7.81$\pm$ 0.04& -1.38$\pm$ 0.07& -0.87$\pm$ 0.11&      \nodata   & -2.11$\pm$ 0.09\\
U1104  & +002 +007 & 13630 $\pm$ $^{1140}_{830}$&  8.08&  7.96&  7.94$\pm$ 0.05& -1.70$\pm$ 0.09& -0.64$\pm$ 0.13& -1.42$\pm$ 0.12& -2.24$\pm$ 0.10\\
U1104  & +004 -001 & 13780 $\pm$ $^{1410}_{980}$&  8.08&  8.03&  7.93$\pm$ 0.05& -1.63$\pm$ 0.11& -0.78$\pm$ 0.14&      \nodata   &    \nodata    \\
U1104  & +004 -001 & 11500 $\pm$ 1500           &  8.10&  8.09&  8.10$\pm$ 0.10& -1.60$\pm$ 0.14&  \nodata       & -1.41$\pm$ 0.25& -2.23$\pm$ 0.17\\
U1175  & -004 +007 & 11200 $\pm$ 1500           &  7.85&  8.17&  7.85$\pm$ 0.10& -1.44$\pm$ 0.17&  \nodata       &      \nodata   &    \nodata    \\
U1175  & -002 -001 & 12200 $\pm$ 1500           &  7.82&  7.99&  7.81$\pm$ 0.10& -1.51$\pm$ 0.13&  \nodata       &      \nodata   &    \nodata    \\
U1175  & +000 -007 & 13500 $\pm$ 1800           &  7.75&  7.78&  7.75$\pm$ 0.10& -1.55$\pm$ 0.15& -0.50$\pm$ 0.25&      \nodata   &    \nodata    \\
U1281  & +008 +013 & 13100 $\pm$ 1500           &  7.78&  7.75&  7.78$\pm$ 0.10& -1.29$\pm$ 0.13&     \nodata    &      \nodata   & -2.14$\pm$ 0.17\\
HKK L14& +002 -004 & 14500 $\pm$ 2000           & 7.65&  7.54&  7.65$\pm$ 0.10& -1.26$\pm$ 0.14& -0.91$\pm$ 0.24&      \nodata   & -2.08$\pm$ 0.16\\
U2023  & -036 -010 & 12500 $\pm$ 1500           &  7.92&  7.91&  7.92$\pm$ 0.10& -1.20$\pm$ 0.13&     \nodata    &      \nodata   & -2.18$\pm$ 0.16\\
U2023  & -029 +023 & 10400 $\pm$ 1200           &  8.15&  8.38&  8.14$\pm$ 0.11& -1.39$\pm$ 0.18&     \nodata    &      \nodata   &    \nodata    \\
U2023  & -028 +023 &  9800 $\pm$ 1000           &  8.20&  8.52&  8.20$\pm$ 0.10& -1.51$\pm$ 0.17&     \nodata    &      \nodata   &    \nodata    \\
U2023  & -020 +023 & 11800 $\pm$ 1500           &  8.00&  8.29&  8.00$\pm$ 0.10& -1.39$\pm$ 0.14&     \nodata    &      \nodata   & -2.17$\pm$ 0.19\\
U2023  & -014 +024 & 12600 $\pm$ 1500           &  7.95&  7.94&  7.95$\pm$ 0.10& -1.30$\pm$ 0.12& -0.90$\pm$ 0.23& -1.62$\pm$ 0.17& -2.14$\pm$ 0.15\\
U2023  & -013 +022 & 11600 $\pm$ 1500           &  8.15&  8.05&  8.15$\pm$ 0.10& -1.33$\pm$ 0.14& -0.70$\pm$ 0.27& -1.62$\pm$ 0.21& -2.26$\pm$ 0.17\\
U2023  & -008 +025 & 12750 $\pm$ $^{410}_{350}$ &  8.03&  7.97&  8.00$\pm$ 0.03& -1.36$\pm$ 0.06& -0.78$\pm$ 0.07& -1.53$\pm$ 0.07& -2.18$\pm$ 0.07\\
U2023  & -006 +022 & 12530 $\pm$ $^{360}_{300}$ &  7.98&  7.84&  8.02$\pm$ 0.03& -1.35$\pm$ 0.06& -0.84$\pm$ 0.07& -1.54$\pm$ 0.07& -2.08$\pm$ 0.08\\
U2023  & +028 +020 & 11500 $\pm$ 1500           &  8.07&  8.24&  8.07$\pm$ 0.10& -1.35$\pm$ 0.14&     \nodata    &      \nodata   &    \nodata    \\
U3647  & -006 -018 & 12140 $\pm$ $^{700}_{540}$ &  7.97&  7.82&  8.07$\pm$ 0.04& -1.28$\pm$ 0.07& -0.88$\pm$ 0.10& -1.64$\pm$ 0.11& -2.10$\pm$ 0.09\\
U3647  & +004 +002 & 12000 $\pm$ 1500           &  8.15&  7.94&  8.14$\pm$ 0.10& -1.25$\pm$ 0.13& -0.83$\pm$ 0.25& -1.46$\pm$ 0.23& -1.96$\pm$ 0.16\\
U3647  & +006 +006 & 13100 $\pm$ 1500           &  7.98&  7.82&  7.97$\pm$ 0.10& -1.21$\pm$ 0.12& -0.91$\pm$ 0.21&      \nodata   & -2.12$\pm$ 0.15\\
U3672  & -052 +091 & 12000 $\pm$ 1500           &  7.78&  8.02&  7.78$\pm$ 0.10&    \nodata    &      \nodata    &      \nodata   &    \nodata    \\
U4117  & -011 +017 & 13100 $\pm$ 1500           &  7.88&  7.80&  7.89$\pm$ 0.10& -1.45$\pm$ 0.14&     \nodata    &      \nodata   &    \nodata    \\
U4117  & -001 +008 & 12200 $\pm$ 1500           &  8.00&  8.04&  8.00$\pm$ 0.10& -1.58$\pm$ 0.14&     \nodata    &      \nodata   &    \nodata    \\
U4117  & +012 +003 & 14500 $\pm$ 2000           &  7.79&  7.65&  7.79$\pm$ 0.10& -1.52$\pm$ 0.16& -0.82$\pm$ 0.24& & -2.16$\pm$ 0.17\\
U4483  & -001 +016 & 16520 $\pm$ $^{590}_{450}$ &  7.62&  7.50&  7.56$\pm$ 0.03& -1.57$\pm$ 0.07& -0.84$\pm$ 0.07& -1.62$\pm$ 0.08& -2.18$\pm$ 0.08\\
U4483  & +001 -006 & 14700 $\pm$ 2000           &  7.70&  7.58&  7.69$\pm$ 0.09& -1.59$\pm$ 0.13& -0.80$\pm$ 0.24&      \nodata   & -2.22$\pm$ 0.17\\
CGCG 007-025& -008 +007 & 14770 $\pm$ $^{530}_{500}$& 7.86&  7.72&  7.83$\pm$ 0.03& -1.55$\pm$ 0.07& -0.83$\pm$ 0.08& -1.47$\pm$ 0.08& -2.33$\pm$ 0.08\\
CGCG 007-025& -004 +000 & 16010 $\pm$ $^{340}_{330}$& 7.89&  7.74&  7.78$\pm$ 0.03& -1.47$\pm$ 0.06& -0.84$\pm$ 0.06& -1.63$\pm$ 0.07& -2.20$\pm$ 0.07\\
CGCG 007-025& +000 -009 & 13920 $\pm$ $^{500}_{430}$& 7.82&  7.70&  7.86$\pm$ 0.03& -1.47$\pm$ 0.07& -0.76$\pm$ 0.08& -1.51$\pm$ 0.08& -2.34$\pm$ 0.08\\
U5288  & -007 +010 & 14320 $\pm$ $^{510}_{500}$ &  8.00&  7.86&  7.90$\pm$ 0.03& -1.26$\pm$ 0.07& -0.71$\pm$ 0.07& -1.51$\pm$ 0.08& -2.12$\pm$ 0.08\\
U5288  & -004 +006 & 14570 $\pm$ $^{840}_{1270}$&  8.12&  8.00&  7.92$\pm$ 0.05& -1.28$\pm$ 0.08& -0.68$\pm$ 0.13& -1.58$\pm$ 0.33& -2.21$\pm$ 0.10\\
U5288  & -000 +000 & 14270 $\pm$ $^{1850}_{1200}$& 8.15&  8.08&  7.90$\pm$ 0.06& -1.28$\pm$ 0.10& -0.84$\pm$ 0.17&      \nodata   & -2.13$\pm$ 0.12\\
U5288  & +006 -013 & 11000 $\pm$ 1500           &  8.30&  8.17&  8.26$\pm$ 0.10& -1.54$\pm$ 0.16&    \nodata     &      \nodata   & -2.24$\pm$ 0.19\\
U5288  & +023 -034 & 12200 $\pm$ 1500           &  8.10&  7.95&  8.09$\pm$ 0.10& -1.76$\pm$ 0.18& -0.87$\pm$ 0.25&      \nodata   & -2.12$\pm$ 0.18\\
U5288  & +026 -033 & 12500 $\pm$ 1500           &  8.02&  7.89&  8.02$\pm$ 0.10& -1.53$\pm$ 0.16&    \nodata     &      \nodata   &    \nodata    \\
UA292  & +004 +013 & 16820 $\pm$ $^{530}_{470}$ &  7.38&  7.22&  7.30$\pm$ 0.03& -1.45$\pm$ 0.07& -0.82$\pm$ 0.07&      \nodata   & -2.32$\pm$ 0.07\\
UA292  & +008 +012 & 15840 $\pm$ $^{920}_{700}$ &  7.40&  7.29&  7.36$\pm$ 0.04& -1.43$\pm$ 0.07& -0.78$\pm$ 0.09&      \nodata   & -2.37$\pm$ 0.08\\
UA292  & +026 +004 & 12200 $\pm$ 1500           &  7.40&  7.64&  7.40$\pm$ 0.10& -1.38$\pm$ 0.13&    \nodata     &      \nodata   &    \nodata    \\
UA292  & +035 -003 & 17230 $\pm$ $^{870}_{710}$ &  7.38&  7.20&  7.31$\pm$ 0.03& -1.37$\pm$ 0.07& -0.73$\pm$ 0.08&      \nodata   & -2.30$\pm$ 0.08\\
UA292  & +038 +000 & 14700 $\pm$ 2000           &  7.39&  7.24&  7.40$\pm$ 0.10& -1.40$\pm$ 0.14&    \nodata     &      \nodata   &    \nodata    \\
U8651  & +045 +036 & 12450 $\pm$ $^{850}_{680}$ &  7.78&  7.73&  7.85$\pm$ 0.04& -1.60$\pm$ 0.09& -0.79$\pm$ 0.12& -1.37$\pm$ 0.12& -2.28$\pm$ 0.10\\
U9240  & +004 -037 & 13350 $\pm$ $^{420}_{330}$ &  7.92&  7.79&  7.95$\pm$ 0.03& -1.60$\pm$ 0.06& -0.75$\pm$ 0.07& -1.60$\pm$ 0.08& -2.15$\pm$ 0.08\\
U9992  & +007 +010 & 11200 $\pm$ 1500           &  7.85&  8.21&  7.84$\pm$ 0.11& -1.29$\pm$ 0.16&    \nodata     &      \nodata   &    \nodata    \\
U9992  & +011 -003 & 12890 $\pm$ $^{5340}_{1800}$& 7.82&  7.76&  7.88$\pm$ 0.12& -1.26$\pm$ 0.19& -0.90$\pm$ 0.34&      \nodata   & -2.23$\pm$ 0.23\\
U9992  & +013 -012 & 11200 $\pm$ 1500           &  8.00&  8.25&  8.00$\pm$ 0.10& -1.39$\pm$ 0.16& -0.86$\pm$ 0.30&      \nodata   & -2.13$\pm$ 0.19\\
U10445 & -034 +053 & 10500 $\pm$ 2000           &  8.30&  8.07&  8.29$\pm$ 0.20& -1.24$\pm$ 0.24&    \nodata     &      \nodata   &    \nodata    \\
U10445 & -026 -002 & 12620 $\pm$ $^{1400}_{930}$&  7.95&  8.01&  7.95$\pm$ 0.06& -1.18$\pm$ 0.09& -0.75$\pm$ 0.16&      \nodata   & -2.13$\pm$ 0.12\\
U10445 & -025 +050 & 11200 $\pm$ 1500           &  7.98&  8.18&  7.97$\pm$ 0.10& -1.25$\pm$ 0.17&    \nodata     &      \nodata   &    \nodata    \\
U10445 & -019 +048 & 12000 $\pm$ 1500           &  8.05&  8.11&  8.05$\pm$ 0.10& -1.26$\pm$ 0.13& -0.66$\pm$ 0.25& -1.52$\pm$ 0.17& -2.27$\pm$ 0.16\\
U10445 & -015 +047 & 12800 $\pm$ $^{2230}_{1280}$& 8.00&  8.05&  7.96$\pm$ 0.07& -1.17$\pm$ 0.12& -0.72$\pm$ 0.22&      \nodata   & -2.22$\pm$ 0.15\\
U10445 & -010 +045 & 11200 $\pm$ 1500           &  8.10&  8.17&  8.10$\pm$ 0.10& -1.33$\pm$ 0.15&    \nodata     &      \nodata   & -2.26$\pm$ 0.19\\
U10445 & -003 +010 & 13900 $\pm$ 2000           &  7.90&  8.11&  7.90$\pm$ 0.10& -1.06$\pm$ 0.15&    \nodata     & -1.42$\pm$ 0.24&    \nodata    \\
U11755 & -003 +003 & 11520 $\pm$ $^{290}_{240}$ &  7.91&  7.81&  8.06$\pm$ 0.03& -1.12$\pm$ 0.06& -0.80$\pm$ 0.07& -1.56$\pm$ 0.07& -2.03$\pm$ 0.07\\
U11755 & -001 +000 & 12400 $\pm$ $^{1380}_{880}$&  7.98&  7.98&  7.99$\pm$ 0.05& -1.08$\pm$ 0.09& -0.62$\pm$ 0.16& -1.53$\pm$ 0.12& -2.14$\pm$ 0.11\\
U12713 & -009 -003 & 13900 $\pm$ 2000           &  7.75&  7.67&  7.75$\pm$ 0.10& -1.60$\pm$ 0.15&    \nodata     &      \nodata   & -2.28$\pm$ 0.18\\
U12713 & -009 -003 & 13900 $\pm$ 2000           &  7.75&  7.67&  7.75$\pm$ 0.10& -1.49$\pm$ 0.16&    \nodata     &      \nodata   & -2.36$\pm$ 0.21\\
U12713 & -001 -003 & 13780 $\pm$ $^{1680}_{1120}$& 7.84&  7.79&  7.80$\pm$ 0.06& -1.54$\pm$ 0.11& -0.81$\pm$ 0.16&       \nodata  & -2.23$\pm$ 0.12\\
U12713 & +004 +004 & 13200 $\pm$ 1500           &  7.75&  7.80&  7.75$\pm$ 0.10& -1.50$\pm$ 0.12&    \nodata     &      \nodata   &    \nodata    \\
\enddata
\end{deluxetable}

\clearpage

\begin{deluxetable}{lcccccccc}
\tabletypesize{\footnotesize}
\tabcolsep 4pt
\tablewidth{0pt}
\tablecaption{Parameters of Additional Dwarf Irregular Galaxies\label{tab:others}}
\tablehead{
\colhead{} & \colhead{Distance}& \colhead{} & \colhead{} & \colhead{}& \colhead{} & \colhead{} & \colhead{12 + } & \colhead{}\\[.2ex]
 \colhead{Galaxy}&\colhead{[Mpc]}& \colhead{M$_{\rm B}$} & \colhead{(B-V)$_0$} & \colhead{M$_{\rm H}$/L$_{\rm B}$} & \colhead{$\mu_B^0$}& \colhead{log(M$_{\rm dyn}$)} & \colhead{log(O/H)} & \colhead{log(N/O)}
}
\startdata
Leo A     &  0.69 & -11.52 & 0.26 $\pm$ 0.10 & 1.3 &\nodata&\nodata& 7.30 $\pm$ 0.05 &-1.52 $\pm$ 0.15
\\
GR 8      &   2.2 & -12.12 & 0.38 $\pm$ 0.10 & 0.8 &\nodata&  7.29 & 7.65 $\pm$ 0.06 &-1.49 $\pm$ 0.07
\\
UGC 9128  &   2.5 & -12.63 & 0.25 $\pm$ 0.05 & 1.1 &  22.8 &  7.90 & 7.75 $\pm$ 0.05 &-1.80 $\pm$ 0.12
\\
UGC 2684  &  5.56 & -13.03 & 0.34 $\pm$ 0.02 & 2.8 &  25.7 &  8.99 & 7.60 $\pm$ 0.10 &-1.38 $\pm$ 0.07
\\
UGC 8024  &   3.2 & -13.33 &     \nodata     & 7.3 &\nodata&  9.42 & 7.67 $\pm$ 0.06 &-1.68 $\pm$ 0.13
\\
UGCA 20   &  8.65 & -14.13 & 0.30 $\pm$ 0.01 & 2.6 &  24.2 &  9.34 & 7.60 $\pm$ 0.10 &-1.56 $\pm$ 0.15
\\
UGC 3174  &  7.86 & -14.76 & 0.41 $\pm$ 0.02 & 2.2 &  23.4 &  9.48 & 7.80 $\pm$ 0.10 &-1.56 $\pm$ 0.15
\\
UGC 521   &  10.9 & -15.17 & 0.30 $\pm$ 0.02 & 2.2 &  20.6 &  9.60 & 7.86 $\pm$ 0.04 &-1.66 $\pm$ 0.12
\\
UGC 5764  &  7.21 & -15.3  &     \nodata     & 0.7 &\nodata&  8.99 & 7.95 $\pm$ 0.04 &-1.34 $\pm$ 0.08
\\
UGCA 357  &  16.5 & -15.50 & 0.41 $\pm$ 0.01 & 3.4 &  25.2 &  9.91 & 8.05 $\pm$ 0.05 &-1.55 $\pm$ 0.11
\\
UGC 891   &  10.5 & -15.53 & 0.46 $\pm$ 0.03 & 1.8 &  23.4 &  9.53 & 8.20 $\pm$ 0.10 &-1.52 $\pm$ 0.13
\\
UGC 5716  &  16.0 & -15.53 & 0.42 $\pm$ 0.01 & 2.5 &  24.1 &  9.89 & 8.10 $\pm$ 0.10 &-1.59 $\pm$ 0.10
\\
UGC 300   &  19.8 & -15.72 & 0.50 $\pm$ 0.01 & 1.4 &  23.3 &\nodata& 7.80 $\pm$ 0.03 &-1.50 $\pm$ 0.08
\\
Haro 43   &  26.5 & -16.31 & 0.41 $\pm$ 0.02 & 2.0 &  20.2 &  9.96 & 8.20 $\pm$ 0.10 &-1.41 $\pm$ 0.10
\\
UGC 5829  &  8.02 & -16.6  &     \nodata     & 1.3 &\nodata&  9.89 & 8.30 $\pm$ 0.10 &-1.76 $\pm$ 0.12
\\
UGC 11820 &  17.9 & -16.94 & 0.37 $\pm$ 0.02 & 2.1 &  23.7 & 10.17 & 8.00 $\pm$ 0.20 &-1.59 $\pm$ 0.10
\\
UGC 191   &  17.6 & -17.52 & 0.46 $\pm$ 0.03 & 0.9 &  22.3 &  9.87 & 8.10 $\pm$ 0.05 &-1.40 $\pm$ 0.10
\\
UGC 634   &  31.4 & -17.67 & 0.39 $\pm$ 0.03 & 2.0 &  23.1 &  9.98 & 8.18 $\pm$ 0.03 &-1.58 $\pm$ 0.08
\\
UGC 2984  &  20.6 & -18.89 & 0.20 $\pm$ 0.02 & 0.5 &  20.5 &  9.99 & 8.30 $\pm$ 0.20 &-1.57 $\pm$ 0.10
\\
\enddata
\tablecomments{Distances are calculated from
systemic velocities using a Virgocentric infall model and
an H$_0$ of 75 km s$^{-1}$ Mpc$^{-1}$ except for
Leo A (Dohm-Palmer et al.\ 1998), GR 8 (Dohm-Palmer et al.\ 1998),
and UGC 9128 (Aparicio et al.\ 2000)
which have TRGB distances in the literature.
Colors, magnitudes, M$_H$/L$_B$, and oxygen and nitrogen
abundances are from van Zee, Haynes, \& Salzer (1997a,b)
and van Zee, Skillman, \& Haynes (2005).
}
\end{deluxetable}

\clearpage

\clearpage

\begin{thebibliography}{}
\bibitem[Alloin et al.(1979)]{ACJV79} Alloin, D., Collin-Souffrin, S., Joly, M., \& Vigroux, L. 1979, \aap, 78, 200
\bibitem[Aparicio et al.(2000)]{ATK00} Aparicio, A., Tikhonov, N., \& Karachentsev, I. 2000, ]aj, 119,
177
\bibitem[Broeils \& Rhee(1997)]{BR97} Broeils, A. H., \& Rhee, M.--H. 1997, \aap, 324, 877
\bibitem[Bruzual \& Charlot(1996)]{BC96} Bruzual, A. G., \& Charlot, S. 1996 in AAS CD-ROM Series, Vol
7, Astrophysics on Disc (Washington: AAS)
\bibitem[Campbell, Terlevich, \& Melnick(1986)]{CTM86} Campbell, A., Terlevich, R. \& Melnick, J. 1986, \mnras, 223, 811
\bibitem[Carigi et al.(1999)]{CCP99} Carigi, L., Col\'{i}n, P., \& Peimbert, M. 1999, \apj, 514, 787
\bibitem[Clayton(1983)]{C83} Clayton, D. 1983, Principles of Stellar Evolution and Nucleosynthesis (Chicago, University of Chicago Press)
\bibitem[De Robertis, Dufour, \& Hunt(1987)]{DDH87} De Robertis, M. M., Dufour,
R. J., \& Hunt, R. W. 1987, \jrasc, 81, 195
\bibitem[Dolphin et al.(2001)]{U4483} Dolphin, A. E., Makarova, L., et al. 2001, \mnras, 324, 249
\bibitem[Dohm-Palmer et al.(1998)]{De98} Dohm-Palmer, R. C., et al. 1998, \aj, 116, 1227
\bibitem[Dopita \& Evans(1986)]{DE86} Dopita, M. A., \& Evans, I. N. 1986, \apj, 307, 431
\bibitem[Edmunds(1990)]{E90} Edmunds, M. G. 1990, \mnras, 246, 678
\bibitem[Edmunds \& Pagel(1978)]{EP78} Edmunds, M. G., \& Pagel, B. E. J. 1978, \mnras, 185, 77p
\bibitem[Edmunds \& Pagel(1984)]{EP84} Edmunds, M. G., \& Pagel, B. E. J. 1984, \mnras, 211, 507
\bibitem[Elmegreen, Elmegreen, \& Morris(1980)]{EEM80}  Elmegreen, B. G., Elmegreen, D. M., \& Morris,
M. 1980, \apj, 240, 455
\bibitem[Ferrara \& Tolstoy(2000)]{FT00} Ferrara, A., \& Tolstoy, E. 2000, \mnras, 313, 291
\bibitem[Gallagher \& Hunter(1984)]{GH84} Gallagher, J. S., \& Hunter, D. A. 1984, \araa, 22, 37
\bibitem[Garnett(1990)]{G90} Garnett, D. R. 1990, \apj, 363, 142
\bibitem[Garnett(2002)]{G02} Garnett, D. R. 2002, \apj, 581, 1019
\bibitem[Henry et al.(2000)]{HEK00} Henry, R. B. C., Edmunds, M. G., \& K\"oppen, J. 2000, \apj, 541, 660
\bibitem[Hidalgo-G\'amez \& Olofsson(1998)]{HGO98} Hidalgo-G\'amez, A. M., \& Olofsson, K. 1998, \aap,
334, 45
\bibitem[Hidalgo-G\'amez \& Olofsson(2002)]{HGO02} Hidalgo-G\'amez, A. M., \& Olofsson, K. 2002, \aap,
389, 836
\bibitem[Howarth(1983)]{H83} Howarth, I. D. 1983, \mnras, 203, 301
\bibitem[Hummer \& Storey(1987)]{HS87} Hummer, D. G., \& Storey, P. J. 1987, \mnras, 224, 801
\bibitem[Hunter \& Hoffman(1999)]{HH99} Hunter, D. A., \& Hoffman, L. 1999, \aj, 117, 2789
\bibitem[Izotov \& Thuan(1998)]{IT98} Izotov, Yu. I., \& Thuan, T. X. \apj, 500, 188
\bibitem[Izotov \& Thuan(1999)]{IT99} Izotov, Yu. I., \& Thuan, T. X. \apj, 511, 639
\bibitem[Izotov \& Thuan(2004)]{IT04} Izotov, Yu. I., \& Thuan, T. X. \apj, 602, 200
\bibitem[Izotov et al.(1994)]{ITL94} Izotov, Yu. I., Thuan, T. X., \& Lipovetsky, V. A. 1994, \apj, 435, 647
\bibitem[Izotov et al.(1997)]{ITL97} Izotov, Yu. I., Thuan, T. X., \& Lipovetsky, V. A. 1997, \apjs, 108, 1
\bibitem[Karachentsev et al.(2002)]{Ke02} Karachentsev, I. D., Sharina, M. E., Makarov, D. I., Dolphin, A. E., Grebel, E. K., Geisler, D., Guhathakurta, P., Hodge, P. W., Karachentseva, V. E., Sarajedini, A., \& Seitzer, P. 2002, \aap, 389, 812
\bibitem[Karachentsev et al.(2004)]{Ke04} Karachentsev, I. D., Karachentseva, V. E., Huchtmeier, W. K., \& Makarov, D. I. 2004, \aj, 127, 2031
\bibitem[Kewley \& Dopita(2002)]{KD02} Kewley, L. J., \& Dopita, M. A. 2002, \apjs, 142, 35
\bibitem[Kennicutt \& Skillman(2001)]{KS01} Kennicutt, R. C., \& Skillman, E. D. 2001, \aj, 121, 1461
\bibitem[Kinman \& Davidson(1981)]{KD81} Kinman, T. D., \& Davidson, K. 1981, \apj, 243, 127
\bibitem[Kniazev et al.(2004)]{KPGLP04} Kniazev, A. Y., Pustilnik, S. A., Grebel, E. K., Lee, H. A., \& Pramskij, A. G. 2004, \apjs, 153, 429
\bibitem[Kobulnicky \& Skillman(1996)]{KS96} Kobulnicky, H. A., \& Skillman, E. D. 1996, \apj, 471, 211
\bibitem[Kobulnicky \& Skillman(1997)]{KS97} Kobulnicky, H. A., \& Skillman, E. D. 1997, \apj, 489, 636
\bibitem[Kobulnicky \& Skillman(1998)]{KS98} Kobulnicky, H. A., \& Skillman, E. D. 1998, \apj, 497, 601
\bibitem[Kobulnicky et al.(1997)]{KSRWR97} Kobulnicky, H. A., Skillman, E. D., Roy, J.--R., Walsh, J. R., \& Rosa, M. R. 1997, \apj, 477, 679
\bibitem[Kobulnicky et al.(2003)]{Ke03} Kobulnicky, H. A., Willmer, C. N. A., Phillips, A. C. et al. 2003, \apj, 499, 1006
\bibitem[K\"oppen \& Hensler(2005)]{KH05} K\"oppen, J. \& Hensler, G. 2005, \aap, 434, 531
\bibitem[Larsen, Sommer-Larsen, \& Pagel(2001)]{LSLP01} Larsen, T. I., Sommer-Larsen, J., \& Pagel, B.
E. J. 2001, \mnras, 323, 555
\bibitem[Lee, Grebel, \& Hodge(2003)]{LGH03} Lee, H., Grebel, E. K., \& Hodge, P. W. 2003, \aap, 401, 141
\bibitem[Lee et al.(2003a)]{LMKRS03} Lee, H., McCall, M. L., Kingsburgh, R. L., \& Stevenson, C. C. 2003a, \aj, 125, 146
\bibitem[Lee et al.(2003b)]{LMR03} Lee, H., McCall, M. L., \& Richer, M. G. 2003b, \aj, 125, 2975
\bibitem[Lee \& Skillman(2004)]{LS04} Lee, H., \& Skillman, E. D. 2004, \apj, 614, 698
\bibitem[Lee, Salzer, \& Melbourne(2004)]{LSM04} Lee, J. C., Salzer, J. J., \& Melbourne, J. 2004, \apj, 616, 752
\bibitem[Lequeux et al.(1979)]{LPRST79} Lequeux, J., Peimbert, M., Rayo, J. F., Serrano, A., \& Torres-Peimbert, S. 1979, \aap, 80, 155
\bibitem[Mac Low \& Ferrara(1999)]{MF99}   Mac Low, M.--M., \& Ferrara, A. 1999, \apj, 513, 142
\bibitem[Maddox et al.(1990)]{APM} Maddox, S. J., Sutherland, W. J., Efstathiou, G., \& Loveday, J.
          1990, \mnras, 243, 692
\bibitem[Ma\'iz-Apell\'aniz et al.(2002)]{MA02} Ma\'iz-Apell\'aniz, J., Cieza, L., MacKenty, J. W. 2002, \aj, 123, 1307
\bibitem[Martin, Kobulnicky, \& Heckman(2002)]{MKH02} Martin, C. L., Kobulnicky, H. A., \& Heckman, T.
M. 2002, \apj, 574, 663
\bibitem[Mateo(1998)]{M98} Mateo, M. L. 1998, \araa, 36, 435
\bibitem[McGaugh(1991)]{M91} McGaugh, S. S. 1991, \apj, 380, 140
\bibitem[McGaugh(1994)]{M94} McGaugh, S. S. 1994, \apj, 426, 135
\bibitem[Meynet \& Maeder(2002)]{MM02} Maynet, G., \& Maeder, A. 2002, \aap, 390, 561
\bibitem[Miller \& Hodge(1996)]{MH96} Miller, B. W., \& Hodge, P. 1996, \apj, 458, 467
\bibitem[Mouhcine \& Contini(2002)]{MC02} Mouhcine, M., \& Contini, T. 2002, \aap, 389, 106
\bibitem[Nilson(1973)]{UGC}   Nilson, P., 1973, Uppsala General Catalog of Galaxies (Uppsala) (UGC)
\bibitem[Oke(1990)]{O90}  Oke, J. B. 1990, \aj, 99, 1621
\bibitem[Olofsson(1997)]{O97} Olofsson, K. 1997, \aap, 321, 29
\bibitem[Osterbrock(1989)]{O89} Osterbrock, D. E. 1989, Astrophysics of Gaseous
Nebulae and Active Galactic Nuclei (University Science Books, Mill Valley)
\bibitem[Pagel et al.(1979)]{PEBCS79} Pagel, B. E. J., Edmunds, M. G., Blackwell, D. E., Chun, M. S.,
      \& Smith, G.  1979, \mnras, 189, 95
\bibitem[Pagel et al.(1992)]{Pe92} Pagel, B. E. J., Simonson, E. A., Terlevich, R. J., \& Edmunds,
        M. G. 1992, \mnras, 255, 325
\bibitem[Peimbert \& Costero(1969)]{PC69} Peimbert, M., \& Costero, R. 1969, Bol. Obs. Tonantzintla y Tacubaya, 5, 3
\bibitem[Pilyugin(2000)]{P00} Pilyugin, L. S. 2000, \aap, 362, 325
\bibitem[Pilyugin(2001)]{P01} Pilyugin, L. S. 2001, \aap, 374, 412
\bibitem[Renzini \& Voli(1981)]{RV81} Renzini, A., \& Voli, M. 1981, \aap, 94, 175
\bibitem[Richer \& McCall(1995)]{RM95} Richer, M. G., \& McCall, M. L. 1995, \apj, 445, 642
\bibitem[Roy \& Kunth(1995)]{RK95} Roy, J.--R., \& Kunth, D. 1995, \aap, 294, 432
\bibitem[Salpeter(1955)]{S55} Salpeter, E. E. 1955, \apj, 121, 161
\bibitem[Searle(1971)]{S71} Searle, L. 1971, \apj, 168, 327
\bibitem[Searle \& Sargent(1972)]{SS72} Searle, L. \& Sargent, W. L. W. 1972, \apj, 173, 25
\bibitem[Seaton(1979)]{S79} Seaton, M. J. 1979, \mnras, 185, 57P
\bibitem[Silich \& Tenorio-Tagle(1998)]{ST98}   Silich, S. A., \& Tenorio--Tagle, G. 1998, \mnras, 299, 249
\bibitem[Skillman(1989)]{S89}  Skillman, E. D. 1989, \apj, 347, 883
\bibitem[Skillman, Bomans, \& Kobulnicky(1997)]{SBK97} Skillman, E. D., Bomans, D. J., \& Kobulnicky, H. A. 1997, \apj, 474, 205
\bibitem[Skillman, C\^ot\'e, \& Miller(2003)]{SCM03} Skillman, E. D., C\^ot\'e, S., \& Miller, B. W. 2003, \aj, 125, 610
\bibitem[Skillman et al.(1989)]{SKH89} Skillman, E. D., Kennicutt, R. C., \& Hodge, P. W. 1989, \apj, 347, 875
\bibitem[Skillman et al.(1996)]{SKSZ96} Skillman, E. D., Kennicutt, R. C., Shields, G. A., \& Zaritsky, D. 1996, \apj, 462, 147
\bibitem[Skillman et al.(1994)]{STKGT94} Skillman, E. D., Terlevich, R. J., Kennicutt, R. C., Garnett,
D. R., \& Terlevich, E. 1994, \apj, 431, 172
\bibitem[Stasi\'nska(1980)]{S80} Stasi\'nska, G. 1980, \aap,  84, 320
\bibitem[Stasi\'nska(1990)]{S90} Stasi\'nska, G. 1990, \aaps, 83, 501
\bibitem[Stasi\'nska \& Leitherer(1996)]{SL96}  Stasi\'nska, G., \& Leitherer, C. 1996, \apjs, 107, 661\bibitem[Swaters et al.(2002)]{SvvS02} Swaters, R. A., van Albada, T. S., van der Hulst, J. M., \& Sancisi, R. 2002, \aap, 390, 829
\bibitem[Talbot \& Arnett(1971)]{TA71} Talbot, R. J. Jr., \& Arnett, W. D. 1971, \apj, 170, 409
\bibitem[Taylor, Kobulnicky, \& Skillman(1998)]{TKS98} Taylor, C. L., Kobulnicky, H. A., \& Skillman, E. D. 1998, \aj, 116, 2746
\bibitem[Tenorio-Tagle(1996)]{TT96} Tenorio-Tagle, G. 1996, \aj, 111, 1641
\bibitem[Thuan et al.(1995)]{TIL95} Thuan, T. X., Izotov, Y. I., \& Lipovetsky, V. A. 1995, \apj, 445,
108
\bibitem[Timmes et al.(1995)]{TWW95} Timmes, F. X., Woosley, S. E., \& Weaver, T. A. 1995, \apjs, 98, 617
\bibitem[Torres--Peimbert et al.(1989)]{TPF89} Torres--Peimbert, S., Peimbert, M., \& Fierro, J. 1989,
\apj, 345, 186
\bibitem[Tremonti et al.(2004)]{T04} Tremonti, C. A., Heckman, T. M., Kauffmann, G. et al. 2004, \apj,
613, 898
\bibitem[van Zee(2000a)]{vZ00a} van Zee, L. 2000a, \apj, 543, L31
\bibitem[van Zee(2000b)]{vZ00} van Zee, L. 2000b, \aj, 119, 2757
\bibitem[van Zee(2001)]{vZ01} van Zee, L. 2001, \aj, 121, 2003
\bibitem[van Zee et al.(1997a)]{vHS97a} van Zee, L., Haynes, M. P., \& Salzer, J. J. 1997a, \aj, 114, 2479
\bibitem[van Zee et al.(1997b)]{vHS97b} van Zee, L., Haynes, M. P., \& Salzer, J. J. 1997b, \aj, 114, 2497
\bibitem[van Zee et al.(1998a)]{vSH98} van Zee, L., Salzer, J. J., \& Haynes, M. P. 1998, \apj, 497, L1                                                                                                       
\bibitem[van Zee et al.(1998b)]{vSHOB98} van Zee, L., Salzer, J. J., Haynes, M. P., O'Donoghue, A. A.,
\& Balonek, T. J. 1998, \aj, 116, 2805
\bibitem[van Zee et al.(2004)]{vSH04} van Zee, L., Skillman, E. D., \& Haynes, M. P. 2004, \aj, 128, 121
\bibitem[van Zee et al.(2005)]{vSH05} van Zee, L., Skillman, E. D., \& Haynes, M. P. 2005, \apj, in press
\bibitem[Verter \& Hodge(1995)]{VH95}   Verter, F., \& Hodge, P. W. 1995, \apj,
446, 616
\bibitem[Vila--Costas \& Edmunds(1993)]{VE93} Vila--Costas, M. B., \& Edmunds, M. G. 1993, \mnras, 265, 199
\bibitem[V\'ilchez \& Iglesias-P\'aramo(1998)]{VP98} V\'ilchez, J. M., \& Iglesias-P\'aramo, J. 1998, \apj, 508, 248
\bibitem[V\'ilchez \& Iglesias-P\'aramo(2003)]{VP03} V\'ilchez, J. M., \& Iglesias-P\'aramo, J. 2003, \apjs, 145, 225
\bibitem[Wilson(1995)]{W95}    Wilson, C.D. 1995, \apjl, 448, L97
\bibitem[Woosley \& Weaver(1995)]{WW95} Woosley, S. E., \& Weaver, T. A. 1995, \apjs, 101, 181
\end{thebibliography}
\end{document}